\documentclass[11pt]{article}

\usepackage{algorithm}
\usepackage[noend]{algpseudocode}
\usepackage{amsmath, amssymb, amsthm}
\usepackage[margin=1in]{geometry}

\usepackage{xcolor,xspace}
\usepackage{thm-restate,thmtools}
\usepackage[symbol]{footmisc}
\usepackage{makecell}
\usepackage{booktabs,nicefrac}
\usepackage{subcaption}

\usepackage[hidelinks]{hyperref}

\usepackage{tikz}
\usetikzlibrary{calc, fit, backgrounds, positioning, shapes.geometric, decorations.pathreplacing}

\usepackage{cleveref}
\usepackage{natbib}

\newcommand{\OPT}{\textnormal{OPT}}
\newcommand{\ALG}{\textnormal{ALG}\xspace}

\newcommand{\dTwoE}{\mathbb E}
\newcommand{\dTwoPr}{\mathbb P}
\newcommand{\dTwoOne}{\mathbf 1}

\newcommand{\Ucore}{U_{\text{core}}}
\newcommand{\Uaux}{U_{\text{aux}}}
\newcommand{\cU}{\mathcal U}

\newcommand{\eps}{\varepsilon}

\newtheorem{theorem}{Theorem}
\newtheorem{lemma}{Lemma}

\newtheorem{proposition}{Proposition}
\newtheorem{remark}{Remark}
\newtheorem{definition}{Definition}[section]

\newtheorem{dettwotheorem}{Theorem}[section]
\newtheorem{dettwolemma}[dettwotheorem]{Lemma}

\title{Consistent Submodular Maximization with Full Foresight\thanks{We would like to thank Kevin Schewior, who raised the question addressed in this work at the Dagstuhl Seminar 25471 ``Online Algorithms beyond Competitive Analysis.''}}
\author{%
  Eric Balkanski\thanks{%
    Columbia University, Email: \texttt{eb3224@columbia.edu}%
  }
  \and
  Paul D\"utting\thanks{%
    Google Research. Emails:
    \texttt{\{duetting,silviol,ashkannorouzi,zadim\}@google.com}
  }
  \and
  Federico Fusco\thanks{%
    Sapienza University of Rome.
    Email: \texttt{federico.fusco@uniroma1.it}.
  }
  \and
  Silvio Lattanzi\footnotemark[3]
  \and
  Ashkan Norouzi-Fard\footnotemark[3]
  \and
  Ola Svensson\thanks{%
    EPFL. Email: \texttt{ola.svensson@epfl.ch}.
    Supported by the Swiss State Secretariat for Education,
    Research and Innovation (SERI) under contract number MB22.00054.%
  }
  \and
  Morteza Zadimoghaddam\footnotemark[3]
}

\date{}

\begin{document}
\pagenumbering{roman}
\maketitle
\thispagestyle{empty}

\begin{abstract}
Online algorithms with bounded recourse face three intertwined obstacles: uncertainty about the future, computational limits, and a limited budget for changes. We ask a fundamental question that separates consistency from the other two: what does solution stability cost when the entire future is known and computation is unlimited?

We answer this question for monotone submodular maximization, where an algorithm must maintain a solution of at most $k$ arrived elements, with a constant number of changes per step.
We determine the optimal approximation thresholds: $3/4$ for deterministic algorithms and $\xi^\star \approx 0.7942$ for randomized algorithms. 
For every $\varepsilon>0$, we get within $\varepsilon$ of these thresholds with worst-case recourse polynomial in $1/\varepsilon$, independent of $k$ and the stream length.
We prove matching impossibility results even for coverage functions.

We attain the sharp bounds by optimally balancing two opposing forces: preparing for future arrivals and protecting the current solution. For randomized algorithms, we use addition robustness against prefixes of the known future, which we analyze tightly via an interval-selection problem. For deterministic algorithms, we introduce certificates, paths in a fractional relaxation that maintain high value, which we round into an update schedule.
\end{abstract}

\clearpage
\tableofcontents
\thispagestyle{empty}

\clearpage
\pagenumbering{arabic}
\section{Introduction}

    Online algorithms with recourse have emerged as a central paradigm for optimization over evolving inputs: they maintain good solutions as the input changes, while limiting how much these solutions change, either at each update or in aggregate over the entire sequence. This paradigm has been developed for a broad range of fundamental problems, including Steiner trees~\citep{ImaseW91,GuGK16}, minimum spanning trees and TSP~\citep{MegowSVW16}, scheduling~\citep{SandersSS09}, bin packing~\citep{FeldkordFGGKRW18}, set cover~\citep{GuptaK0P17}, bipartite matching~\citep{BosekLSZ14,BernsteinHR19}, $k$-clustering~\citep{LattanziV17,FichtenbergerLN21}, and submodular maximization~\citep{DuettingFLNZ24,DuttingFLNSZ25}.

    Such algorithms face three intertwined obstacles: uncertainty about the future, computational limits, and a limited budget for changes. Approximation guarantees reflect the joint effect of all three, so they do not reveal what stability itself costs. Isolating this cost requires removing the other two obstacles, which leads to the following question:
    \begin{quote}
        \emph{How much solution quality must be sacrificed to maintain stability when the entire future is known and computation is unlimited?}
    \end{quote}

    We study this question for monotone submodular maximization under a cardinality constraint \citep{NemhauserWF78,KrauseG14}. We give the algorithm full knowledge of the arrival sequence and unrestricted computational power. Each maintained solution must use only elements that have already arrived, and consecutive solutions may differ by only a constant number of elements. Even with full foresight, preparing a solution for future arrivals can compromise its current value, while postponing changes can leave too little time to complete them. The challenge is to plan solutions and transitions that remain close to the current optimum while respecting the recourse bound.

\subsection{The Model}

    Let $f:2^X\to\mathbb{R}_{\geq0}$ be a monotone submodular function accessible through value queries, and let $k\geq1$ be an integer cardinality bound. Elements $x_1,\ldots,x_n\in X$ arrive one at a time, and the algorithm knows the entire arrival sequence in advance.  After the $t$-th arrival, the algorithm must maintain a solution $\ALG_t\subseteq X_t$, where $X_t=\{x_1,\ldots,x_t\}$, satisfying $|\ALG_t|\le k$ for every realization of its randomness. We set $\ALG_0=\varnothing$ and let $\OPT_t$ denote a set maximizing $f(S)$ over all $S\subseteq X_t$ with $|S|\le k$. Computation and the number of value queries are unrestricted, and queries may involve elements that have not yet arrived.

    We evaluate a possibly randomized algorithm according to two guarantees:
    \begin{itemize}
    \item \textbf{Approximation.}
    For $\alpha \leq 1$, an algorithm is an \emph{$\alpha$-approximation} if, on every instance and at every time $t$,
    $
    \mathbb{E}[f(\ALG_t)]\ge \alpha f(\OPT_t).
    $
    The expectation is over the algorithm's internal randomness, with the instance and time $t$ fixed; the guarantee need not hold simultaneously at all times for a single realization. For deterministic algorithms, we use $f(\ALG_t)$ instead of $\mathbb{E}[f(\ALG_t)]$.

    \item \textbf{Consistency.}
    An algorithm is \emph{$C$-consistent} if, on every instance and for every realization of its randomness,
    $
    |\ALG_t\triangle\ALG_{t-1}|\le C
    ~\text{for all }t=1,\ldots,n.
    $
    Here $\triangle$ denotes the symmetric difference.\footnote{$A\triangle B=(A\setminus B)\cup(B\setminus A)$.}
    An algorithm is \emph{consistent} if it is $C$-consistent for some constant $C$ independent of the input instance, including $k$ and $n$. For algorithms parameterized by an accuracy $\varepsilon>0$, the constant $C$ may depend on $\varepsilon$.
    \end{itemize}

    We allow arbitrary constant recourse budgets to identify approximation barriers that persist for every fixed number of changes permitted per arrival.
    
    Apart from the algorithm's knowledge of the arrival sequence, this is the model of consistent submodular maximization introduced by \citet{DuettingFLNZ24} and further studied by \citet{DuttingFLNSZ25} and \citet{FuZT26}, in which each element is revealed only upon arrival.\footnote{These works bound only the number of insertions, $|\ALG_t\setminus\ALG_{t-1}|$. For monotone $f$, this changes recourse bounds by at most a factor of two, since postponing each deletion until an insertion requires space never decreases the value of the solution.} We refer to their setting as \emph{non-clairvoyant} and to ours as \emph{clairvoyant}.

\begin{figure}[t]
\begin{subfigure}{0.5\textwidth}\centering
\scalebox{0.8}{
\begin{tabular}{@{}l|cc@{}}
\toprule
     & poly time &  info-theoretic\\
\midrule
det. & \makecell{$[0.3818,\nicefrac{1}{2}]$\\{\small \citet{DuettingFLNZ24}}} & \makecell{$[0.3818,\nicefrac{1}{2}]$\\{\small \citet{DuettingFLNZ24}}} \\
\midrule
rand. & \makecell{$2 - \sqrt 2 \approx 0.5858$ \\ {\small \citet{FuZT26}}} & \makecell{$\nicefrac{{3}}{{4}}$ {\small (coverage)} and $\nicefrac{{2}}{{3}}$ {\small (submod.)}\\ {\small \citet{DuttingFLNSZ25}}}\\
\bottomrule
\end{tabular}
} 
\caption{Non-clairvoyant algorithms}
\label{tab:overview_non_clairvoyant}
\end{subfigure}
\hfill
\begin{subfigure}{0.4\textwidth}\centering
\scalebox{0.8}{
\begin{tabular}{@{}l|cc@{}}
\toprule
     & poly time &  info-theoretic\\
\midrule
det. & \makecell{$[0.3818,1-\nicefrac{1}{e}]$\\{\small \citet{DuettingFLNZ24}}} & \makecell{$\nicefrac{3}{4}$\\{\small Thm.~\ref{thm:det_ub} \& Thm.~\ref{thm:positive_deterministic}}}  \\
\midrule
rand. & \makecell{$[2-\sqrt 2,1-\nicefrac{1}{e}]$ \\ {\small \citet{FuZT26}}} & \makecell{ $\xi^\star\approx 0.7942$ \\ {\small Thm.~\ref{thm:rand_ub_overview} \& Thm.~\ref{thm:positive_randomized}}}\\
\bottomrule
\end{tabular}
} 
\caption{Clairvoyant algorithms}
\label{tab:overview_clairvoyant}
\end{subfigure}
\caption{Overview of results. A single value is a tight threshold; an interval gives the best known lower and upper bounds. \emph{Poly time} refers to algorithms that make polynomially many value queries and run in polynomial time, and \emph{info-theoretic} to algorithms with unrestricted computation. The upper bound of $1-1/e$ in polynomial time holds even without consistency constraints~\citep{NemhauserW78,Feige98}.}
\label{fig:overview}
\end{figure}

    \subsection{Results and Techniques}

    We establish tight information-theoretic bounds for consistent clairvoyant algorithms, both deterministic and randomized.

    For randomized algorithms we show:

    \medskip\noindent
    \textbf{Main Result 1 (Tight randomized bound; see Thm.~\ref{thm:rand_ub_overview} \& Thm.~\ref{thm:positive_randomized}).}
    \emph{For every $\varepsilon\in(0,1)$, there is a randomized clairvoyant algorithm for monotone submodular maximization under a cardinality constraint that is $O(\varepsilon^{-2})$-consistent and achieves a $(\xi^\star-\varepsilon)$-approximation, where $\xi^\star\approx0.7942$.\footnote{The constant $\xi^\star$ is the unique root in $(1/2,1)$ of $89\xi^4-44\xi^3-19\xi^2-3\xi+1=0$.} No consistent randomized clairvoyant algorithm can achieve an approximation ratio greater than $\xi^\star$, even for coverage functions.}
    \medskip
    
    For this result, we handle transitions through the addition-robustness framework of \citet{DuttingFLNSZ25}. The algorithm periodically samples a new set of elements that have already arrived and transitions to it at a random time, so that any fixed time is rarely affected. This framework reduces the problem to constructing an \emph{addition-robust} distribution, whose samples stay competitive in expectation as new elements arrive. 

    Foresight changes the nature of this task: the upcoming arrivals are known, but a single distribution must still be good at all times. Via minimax duality, as in the non-clairvoyant setting, it suffices to compare a single solution, fixed in advance, with the best choice in hindsight when the evaluation time is random. Our key insight is that, because the future arrival prefixes form a known nested sequence, this question can be cast as an \emph{interval-selection problem}, quite different from the correlation-gap question that arises without foresight~\citep{DuttingFLNSZ25}.

    We solve this problem exactly. The crux is to show that the worst case consists of only three consecutive intervals, corresponding to just two future insertions; optimizing over such instances yields $\xi^\star$. The same analysis yields both bounds at once: every instance admits a robust distribution that is $\xi^\star$-competitive, and the worst case translates directly into a family of coverage instances that rules out a better approximation ratio for consistent algorithms.

    For deterministic algorithms we show:

    \medskip\noindent
    \textbf{Main Result 2 (Tight deterministic bound; see Thm.~\ref{thm:det_ub} \& Thm.~\ref{thm:positive_deterministic}).}
    \emph{For every $\varepsilon\in(0,1)$, there is a deterministic clairvoyant algorithm for monotone submodular maximization under a cardinality constraint that is $O(\varepsilon^{-3}\log(1/\varepsilon))$-consistent and achieves a $(3/4-\varepsilon)$-approximation. No consistent deterministic clairvoyant algorithm can achieve an approximation ratio greater than $3/4$, even for coverage functions.}
    \medskip

    The randomized construction places transitions at random times, so that any fixed time is affected by a transition only with small probability. A deterministic algorithm must maintain its approximation guarantee throughout each transition. 
    A simple coverage construction makes this obstacle concrete: the algorithm is forced to move gradually from the rows of a $k\times k$ grid to increasingly valuable columns. Halfway through, its $k/2$ rows and $k/2$ columns cover only $3/4$ of the grid, while the optimal solution covers all of it.
    
    Obtaining the matching algorithmic guarantee requires a way to construct safe transitions. We introduce a framework of \emph{clairvoyant certificates}, which combines a fractional solution that remains valuable throughout a window of arrivals with a transition path of controlled complexity connecting the initial and final optimizers through that solution. The entry and exit portions satisfy the required value guarantees under the initial and final objectives, respectively.
    A certificate thus balances preparing for future arrivals against protecting the current solution.
    The main technical contribution is a reduction that rounds the entire transition path simultaneously, preserving the required value guarantees while controlling the number of changes. The rounding argument establishes the existence of a single realization satisfying all requirements, which can be found deterministically by exhaustive search. Scheduling the resulting changes then yields a consistent algorithm with an arbitrarily small loss in approximation.
    
    This framework reduces the algorithmic problem to establishing the existence of suitable fractional certificates. We give an elementary construction at level $1/\sqrt2 > 1-1/e$ that illustrates the framework, and prove in Lean that every window admits a certificate at the sharp level $3/4$, yielding the optimal deterministic guarantee.

    \paragraph{Discussion.} Even with full foresight and unrestricted computation, constant recourse imposes a strict approximation loss. Our results quantify this loss exactly and show that randomization remains beneficial in the absence of uncertainty about future arrivals. Figure~\ref{fig:overview} compares our thresholds with those for non-clairvoyant algorithms: foresight raises the deterministic threshold from at most $1/2$~\citep{DuettingFLNZ24} to $3/4$, and the randomized threshold from $2/3$ to $\xi^\star$ for submodular functions and from $3/4$ to $\xi^\star$ for coverage functions~\citep{DuttingFLNSZ25}. Without foresight, randomized algorithms do better on coverage functions than on general submodular functions; with foresight, the two classes have the same thresholds.

    \paragraph{Open Problems.} 
    Both of our thresholds exceed $1-1/e$, the best ratio achievable efficiently even without stability constraints~\citep{NemhauserW78,Feige98}. Can efficient clairvoyant algorithms with constant recourse approach $1-1/e$, so that stability costs nothing beyond computation? This would require exploiting foresight: without it, deterministic algorithms cannot exceed $1/2$~\citep{DuettingFLNZ24}, and efficient randomized algorithms cannot exceed $2-\sqrt2$~\citep{FuZT26}. The information-theoretic threshold for deterministic non-clairvoyant algorithms also remains open, between $0.3818$ and $1/2$~\citep{DuettingFLNZ24}. More broadly, it would be interesting to determine the cost of stability for the other problems and constraints studied with recourse.

    \subsection{Related Work}

    \paragraph{Consistent Submodular Maximization.}
    In the non-clairvoyant setting, \citet{DuettingFLNZ24} gave a deterministic $(0.3818-\varepsilon)$-approximation with constant recourse and showed that no consistent deterministic algorithm can exceed $1/2$. \citet{DuttingFLNSZ25} introduced addition robustness and a randomized checkpointing reduction, and determined the information-theoretic thresholds for randomized algorithms: $2/3$ for monotone submodular functions and $3/4$ for coverage functions. \citet{FuZT26} showed that $2-\sqrt2$ is the best ratio achievable with polynomially many value queries. \citet{DuttingFLNSZ26} studied consistency in the fully dynamic setting, with both insertions and deletions. A different notion of stability underlies the submodular objectives chasing problem of \citet{BuchbinderNW25}, in which both the objective and the ground set change over time: their algorithms maintain a $(1-1/e-\varepsilon)$-approximation while keeping the total recourse competitive with that of exact offline algorithms. In contrast, we bound the recourse of every step by a constant.

    \paragraph{Incremental and Multistage Optimization.}
    Several models study how solutions evolve when the input is known in advance. In incremental maximization, the solutions for increasing budgets must be nested, and each must be competitive for its own budget~\citep{BernsteinDGH22,DisserKSW23}; for submodular functions under increasing cardinality constraints, the greedy algorithm achieves $1-1/e$, and \citet{BienkowskiBBCDL26} recently improved on it. In incremental-decremental maximization, an initial solution is transformed one element at a time, and the order must keep every intermediate stage competitive with the optimum for that stage~\citep{DisserKLS26}. In multistage optimization, a sequence of instances is given, and the quality of each stage's solution is traded off against the cost of changing it, or the changes between consecutive stages are bounded~\citep{GuptaTW14,EisenstatMS14,AnNS17,BampisELP18,FluschnikNRZ22}. In our model, by contrast, the cardinality is fixed while the ground set grows, every solution must approximate the current optimum, and only a constant number of elements may change per arrival.

    \paragraph{Streaming, Preemptive, and Dynamic Submodular Maximization.}
    Streaming submodular maximization \citep{ChakrabartiK15, BadanidiyuruMKK14, FeldmanNSZ20} focuses on bounding working memory ($O(k \log k / \eps)$ for a $(\nicefrac{1}{2}-\eps)$-approximation) rather than solution stability. \citet{BuchbinderFS19, ChanEtAl18} study online submodular maximization under cardinality constraints with preemption but no recall.  Fully dynamic submodular maximization under insertions and deletions \citep{LattanziMNTZ20, Monemizadeh20, ChenP22, DuettingFLNZ23, BanihashemBGHJM23} optimizes query update time; however, dynamic algorithms can achieve polylogarithmic \emph{expected amortized} update time while still incurring $\Omega(k)$ worst-case recourse during rebuild steps.

    \paragraph{Online Algorithms with Bounded Recourse.}
    Maintaining solutions with bounded recourse has been widely studied for combinatorial problems including Steiner trees \citep{ImaseW91}, set cover \citep{GuptaK0P17}, $k$-clustering \citep{LattanziV17, FichtenbergerLN21}, and matching \citep{BernsteinHR19}. Studying clairvoyant (offline dynamic) counterparts of online recourse problems separates the intrinsic cost of smooth transitions from online uncertainty; our results show that for submodular maximization, step-by-step consistency imposes a fundamental approximation barrier even under full foresight.
\section{The Tight Randomized Bound}
       \label{sec:sharp-randomized}

The optimal approximation ratio for randomized clairvoyant algorithms with a consistency bound independent of the cardinality constraint is $\xi^\star\approx 0.794$ for both coverage and monotone submodular functions, where $\xi^\star$ is the unique\footnote{Existence and uniqueness of $\xi^\star$ are established in Appendix~\ref{app:unique}.} root in $(\nicefrac 12,1)$ of the polynomial
\[
    \Psi(\xi)=89\xi^4-44\xi^3-19\xi^2-3\xi+1.
\]
We give the impossibility result in Section~\ref{sec:rand-impossibility} using a simple hard instance, then obtain a matching algorithmic guarantee in Sections~\ref{sec:rand-algorithm}--\ref{sec:rand-hard-structure} by showing that this instance is a worst-case instance. The proofs for this section are deferred to Appendix~\ref{app:sharp-randomized}.

\subsection{The Impossibility Result for Randomized Algorithms}
\label{sec:rand-impossibility}

    Quite surprisingly, there is a very simple hard-instance construction based on \emph{only two insertions} which yields this impossibility result for possibly randomized consistent algorithms. Moreover, the instance is based on a coverage function, thus showing---together with the matching algorithm---that there is no separation between coverage and monotone submodular functions. We state the result and provide a proof sketch, while the details can be found in \Cref{app:rand-lower-bound}.

    \begin{restatable}{theorem}{hardrand}
    \label{thm:rand_ub_overview}
        For any $\eps \in [\nicefrac 1k,1]$,
        no randomized $\eps k$-consistent clairvoyant algorithm can achieve
        an approximation ratio better than
        $\xi^\star+O(\eps)\approx 0.7942+O(\eps)$
        for Maximum Coverage subject to a cardinality
        constraint $k$.
    \end{restatable}
    \begin{proof}[Proof sketch]
        We describe a weighted coverage instance on which every randomized
        $\eps k$-consistent clairvoyant algorithm has approximation ratio at most
        $\xi^\star+O(\eps)$. Appendix~\ref{app:rand-lower-bound} converts it to an
        unweighted instance. Partition a universe $U$ of $3k$ elements into
        groups $U_0,U_1,U_2$ of $k$ elements each, with respective
        element weights $1 > u_1 > u_2 > 0$. The arrival sequence consists of the $3k$ singleton
        subsets of $U$, followed by the two large sets $U_0$ and $U_1$, in that
        order. Thus the sequence contains $3k+2$ sets; the singleton sets
        may arrive in any order. Figure~\ref{fig:randomized-hard-instance}
        illustrates the construction.

        \colorlet{elemcol}{red!75!black}
\colorlet{ucol}{blue!70!black}
\colorlet{vcol}{violet!75!black}

\pgfmathsetmacro{\wa}{0.5508}
\pgfmathsetmacro{\wb}{0.4705}
\def\unitH{2.6}
\def\rectW{2.4}

\begin{figure}[t!]
\centering
\resizebox{\linewidth}{!}{%
\begin{tikzpicture}[
  every node/.style = {font=\Huge},
  vset/.style = {
    rectangle,
    rounded corners=2pt,
    draw=vcol,
    line width=1.8pt,
    inner sep=2mm
  },
  vlab/.style = {text=vcol, anchor=south},
]

\newcommand{\Uset}[4]{%
  \node[
    draw=ucol,
    fill=ucol!8,
    line width=1.2pt,
    inner sep=0pt,
    minimum width=\rectW cm,
    minimum height=#3*\unitH cm,
    text=elemcol
  ] (#1) at (#2,0) {$#4$};
}

\Uset{U1}{0}{1}{1}
\Uset{U2}{2.4}{1}{1}
\node at (4.8,0) {$\cdots$};
\Uset{Uk}{7.2}{1}{1}

\node[vset, fit=(U1)(Uk)] (V1) {};
\node[vlab] at ([yshift=1mm]V1.north) {$U_0$};

\Uset{Uk1}{11.0}{\wa}{u_1}
\Uset{Uk2}{13.4}{\wa}{u_1}
\node at (15.8,0) {$\cdots$};
\Uset{U2k}{18.2}{\wa}{u_1}

\node[vset, fit=(Uk1)(U2k)] (V2) {};
\node[vlab] at ([yshift=1mm]V2.north) {$U_1$};

\Uset{U2k1}{22.0}{\wb}{u_2}
\Uset{U2k2}{24.4}{\wb}{u_2}
\node at (26.8,0) {$\cdots$};
\Uset{U3k}{29.2}{\wb}{u_2}

\end{tikzpicture}
}
\caption{The hard coverage instance for the randomized lower bound.}
\label{fig:randomized-hard-instance}
\end{figure}
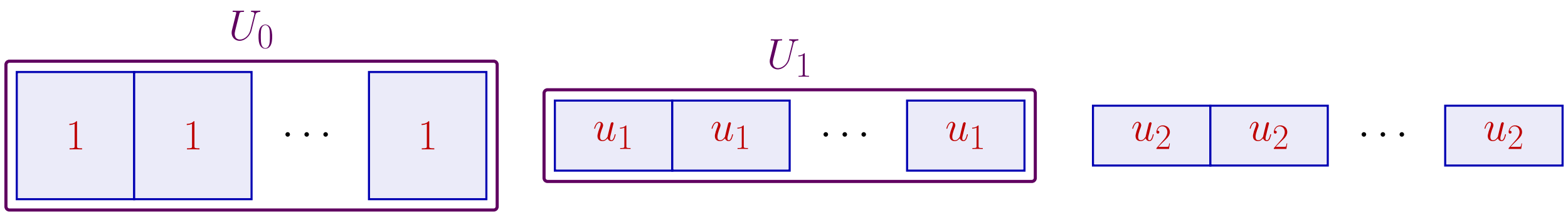

        Relabel the last three time steps as $0$, $1$, and $2$: immediately before
        $U_0$ arrives, immediately after $U_0$ arrives, and immediately after
        $U_1$ arrives. Let $k_i$ be the expected number
        of singleton sets from $U_i$ in the algorithm's solution at time $0$. These counts suffice to
        bound the algorithm's expected value at all three time steps: we  include each large set
        in the algorithm's solution as soon as it arrives, without counting it
        toward the cardinality constraint, and consistency limits the algorithm to at most
        $\eps k$ additional singleton sets per insertion. Since every
        singleton has weight at most $1$, these additions contribute only
        $O(\eps k)$ to its value.

        In this sketch, we omit the \(O(\varepsilon k)\) contribution of these additional singleton sets to the value of the solutions of the algorithm and assume the large sets also do not count toward the cardinality constraint for the optimal solutions. With these simplifications, the algorithm's expected values at time steps $0$, $1$, and $2$ are at most
        $k_0+u_1k_1+u_2k_2$, $k+u_1k_1+u_2k_2$, and $k(1+u_1)+u_2k_2$. The corresponding optimal values are $k$, $k(1+u_1)$, and $k(1+u_1+u_2)$. We can thus bound the optimal approximation ratio with the following LP. 
            \begin{align*}
                \text{Maximize}\quad &\rho\\
                \text{subject to}\quad
                &k_0+k_1+k_2 \leq k,\\
                &k_0+u_1k_1+u_2k_2  \ge \rho \cdot k,\\
                &k+u_1k_1+u_2k_2
                    \geq\rho\left(1+u_1\right)\cdot k, \\
                &k(1+u_1)+u_2k_2
                    \geq\rho\left(1+u_1+u_2\right)\cdot k, \\
                &k_i\geq 0\quad\text{for }i=0,1,2.
            \end{align*}
        To bound the optimal value of this program, we take a convex combination
        of its three approximation constraints with weights $u_2, \frac{u_2}{u_1}-u_2,$ and $1-\frac{u_2}{u_1}$ and obtain
        \[
            \rho \le \frac{1+u_1-u_2}{\left(1+u_1+u_2-u_1u_2-\frac{u_2^2}{u_1}\right)}.
        \]
        To minimize this  bound on $\rho$, we choose $u_1$ as the unique
        root in $(0,1)$ of the polynomial
        \(
            \Phi(x)=x^4+x^3-2x^2-3x+2
        \)
        and set $u_2=u_1^2(1+u_1)$, which makes the bound equal to $1/(1+u_1u_2)=\xi^\star$.
    \end{proof}

\subsection{Overview of the Randomized Algorithm}
\label{sec:rand-algorithm}

\begin{figure}[t!]
\centering
\begingroup
\newcommand{\roadbox}[5]{%
  \node[roadbox, minimum width=#3] (#1) at (#2,0) {};
  \node[font=\normalsize, text=black, align=center, inner sep=0pt]
    at ($(#1.center)+(0,7pt)$) {#4};
  \node[font=\small, text=black, inner sep=0pt]
    at ($(#1.south)+(0,11pt)$) {#5};}
\newcommand{\roadarrow}[3]{%
  \draw[->, >=stealth, line width=0.55pt, black!70,
        shorten <=2pt, shorten >=2pt]
    (#1.east) -- node[above=4pt, font=\normalsize,
                     text=black, inner sep=0pt]
      {Lemma~\ref{#3}} (#2.west);}
\begin{tikzpicture}[x=1pt, y=1pt,
  roadbox/.style={draw=black!55, fill=white, line width=0.45pt,
                  rounded corners=1.5pt, minimum height=82pt,
                  inner sep=0pt, outer sep=0pt}]
\roadbox{consistent}{38.5}{77pt}
  {Consistent\\submodular\\maximization}
  {Theorem~\ref{thm:positive_randomized}}
\roadbox{robust}{177}{84pt}
  {Clairvoyant\\addition-robust\\submodular\\maximization}
  {Section~\ref{sec:rand-robustness}}
\roadbox{interval}{304}{54pt}
  {Interval\\selection}
  {Section~\ref{sec:rand-interval-selection}}
\roadbox{chain}{427}{76pt}
  {Consecutive\\intervals,\\equalizing\\distribution}
  {Section~\ref{sec:rand-hard-structure}}
\roadarrow{consistent}{robust}{thm:meta}
\roadarrow{robust}{interval}{lem:rand-interval-reduction}
\roadarrow{interval}{chain}{lem:rand-chain-reduction}
\end{tikzpicture}
\endgroup
\caption{Proof roadmap for the randomized algorithm.}
\label{fig:rand-reductions}
\end{figure}
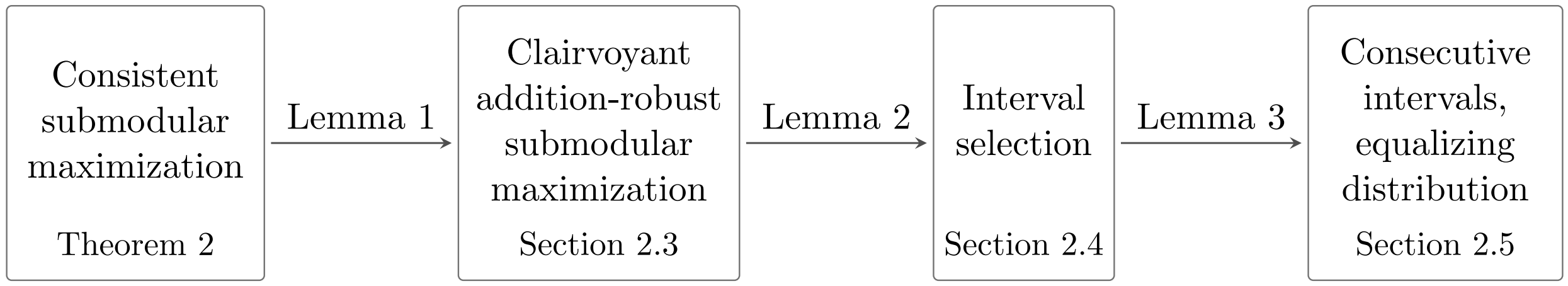

In this section, we present the randomized algorithm that achieves the tight $\xi^\star$ approximation. The result follows from a chain of reductions, summarized in \Cref{fig:rand-reductions}. As a first step, in \Cref{sec:rand-robustness}, we introduce the notion of \emph{clairvoyant} addition-robust algorithms, building on the non-clairvoyant notion introduced by \citet{DuttingFLNSZ25}. Using a similar checkpoint-based construction, we show that finding a solution robust to prefixes of the known arrival sequence suffices to obtain consistent algorithms with a comparable approximation guarantee. Minimax duality applied to the linear program induced by addition-robustness then reformulates the task: instead of finding a single randomized solution that performs well at every prefix, it suffices to find, for each fixed distribution over prefixes, a deterministic solution that performs well in expectation.

Our main technical contribution in this section is to connect this minimax problem to an \emph{interval selection} problem and give a tight analysis of it. In this formulation, each interval corresponds to a prefix of the arrival sequence: its left endpoint is the value of that prefix, and its length is the maximum marginal contribution of a feasible solution to that prefix. In \Cref{sec:rand-hard-structure}, we then reduce the general interval selection problem to instances with consecutive intervals of nonincreasing lengths. Finally, we pinpoint the exact approximation guarantee $\xi^\star$ by explicitly solving the interval selection problem for this restricted family, whose worst-case instance consists of only three intervals and corresponds to the hard coverage instance from Theorem~\ref{thm:rand_ub_overview}.

\begin{restatable}{theorem}{positiverand}
\label{thm:positive_randomized}
    For every $\varepsilon\in(0,1)$, there is a randomized clairvoyant
    algorithm for monotone submodular maximization under a cardinality
    constraint $k$ that is $O(\nicefrac 1{\varepsilon^{2}})$-consistent and achieves a $(\xi^\star-\varepsilon)$-approximation.
    \end{restatable}

\subsection{The Clairvoyant Addition-Robust Problem}
\label{sec:rand-robustness}

    In \emph{clairvoyant addition-robust submodular maximization}, we are given a monotone submodular function $f$, a partition of the ground set into present elements $V_{\mathrm{now}}$ and future elements
$V_{\mathrm{future}}=(v_1,\ldots,v_m)$ whose arrival order is given
in advance, and a cardinality budget $\kappa$.  Write $R_j=\{v_1,\ldots,v_j\}$, with $R_0=\varnothing$, and let
\begin{equation*}
    O_j\in\operatorname*{arg\,max}_{\substack{O\subseteq V_{\mathrm{now}}\\
                                           |O|\leq\kappa}}f(O\cup R_j).
\end{equation*}
 The goal is to find a (possibly randomized) set $S \subseteq V_{\mathrm{now}}$ of cardinality at most $\kappa$ which is ``robust'' with respect to the insertion of all possible prefixes of $V_{\mathrm{future}}$, i.e., such that the value of $S\cup R_j$ is a good approximation of that of $O_j \cup R_j$, for all $j$ at the same time.

\begin{definition}[Clairvoyant Addition Robustness]
 In the clairvoyant setting, a possibly randomized set $S \subseteq V_{\mathrm{now}}$ is $\alpha$-addition-robust if
\begin{equation*}
    \mathbb{E}[f(S\cup R_j)]\geq\alpha f(O_j\cup R_j)
    \qquad \forall j=0,\ldots,m.
\end{equation*}
We call such a distribution $\alpha$-addition-robust.
\end{definition}

The algorithm knows $V_{\mathrm{future}}$, so it can compute up front
all sets $O_i$, which are best responses to the corresponding prefixes
$R_i$. In the consistent problem, however, it cannot move freely from
one $O_i$ to $O_{i+1}$. The addition-robust algorithm therefore commits,
possibly in a randomized way, to one fixed set of present elements. We highlight that the difference with respect to the non-clairvoyant definition in \citet{DuttingFLNSZ25} lies in the fact that here the robustness is only with respect to known prefixes of the known sequence $V_{\mathrm{future}}$ of ``future'' elements, and not with respect to all possible unknown futures. The benchmark $f(O_i\cup R_i)$ upper-bounds the optimum attainable from
$V_{\mathrm{now}}\cup R_i$ with at most $\kappa$ elements. The following reduction
converts addition robustness into a consistent algorithm.

\begin{restatable}{lemma}{metatheorem}
\label{thm:meta}
Suppose that, for every instance of the clairvoyant addition-robust
problem, an algorithm returns an $\alpha$-addition-robust distribution,
where $0\leq\alpha\leq1$.
For every $\gamma\in(0,\nicefrac 18)$, this yields a randomized clairvoyant
$O(\gamma^{-2})$-consistent algorithm that achieves an $(\alpha-3\gamma)$-approximation.
\end{restatable}

The structure of the reduction is similar to the one in \citet{DuttingFLNSZ25}. It relies on a partition of the arrivals into contiguous blocks of $\gamma k$ elements, interleaved by checkpoints. At each checkpoint, the algorithm computes an addition-robust solution using as $V_{\mathrm{now}}$ the elements that have arrived so far, and as $V_{\mathrm{future}}$ all the elements in the following two blocks. Then the algorithm selects uniformly at random a subblock of size $\gamma^2 k$ where it transitions from the addition-robust solution inherited from the previous checkpoint to the solution computed at the current checkpoint. Robustness guarantees that the approximation factor is maintained throughout each block, while the random choice of the transition subblock ensures that, at each time step, the probability of hosting a transition is at most $\gamma$. Consistency is ensured because we use $\gamma^2k$ time steps to transition from one solution to the other maintaining $O(\gamma^{-2})$ consistency. For more details on the checkpoint algorithm and the proof of~\Cref{thm:meta}, we refer to~\Cref{app:rand-checkpoints}.

\paragraph{The addition-robustness LP.} To obtain the desired guarantee, it suffices to consider distributions supported on \(O_0,\ldots,O_m\). The best such distribution is given by the following LP, which generalizes the LP in the analysis of the hard instance for \Cref{thm:rand_ub_overview}.

\begin{align}
                \text{Maximize} \quad & \rho \label{lp:randomized} \\
                \text{subject to} \quad & \sum_{i=0}^m \alpha_i f(O_i \cup R_j) \ge \rho f(O_j\cup R_j)\quad \text{ for } j=0,\dots, m,\notag\\
                & \sum_{i=0}^m \alpha_i = 1, \qquad \alpha_i \ge 0 \quad \text{ for } i=0,\dots, m,\qquad 0\leq\rho\leq1.\notag
            \end{align}

For the normalized formulation below, assume $\kappa\geq1$ and
$f(O_0)>0$. Optimality and monotonicity
then give $f(O_j\cup R_j)\geq f(O_0)>0$ for every $j$. Let $\Delta$ denote the simplex of probability distributions
over $\{0,\ldots,m\}$. Writing $\rho^\star$ for the value of the
linear program~\eqref{lp:randomized}, we have

\[
    \rho^\star = \max_{\alpha \in \Delta} \min_{j=0,\ldots,m}
    \sum_{i=0}^m \alpha_i \cdot \frac{f(O_i \cup R_j)}{f(O_j\cup R_j)}.
\]

This program can be interpreted as a game between the algorithm which picks a distribution over sets $O_i$, knowing that there is an adversary which is going to select the prefix $R_j$ for which the expected normalized value of the  randomized solution is worst. By finite minimax duality, we may reverse the order of play: the adversary first commits to a distribution $\beta\in\Delta$ over prefixes, and the algorithm then chooses a deterministic solution $O_i$ maximizing its expected normalized value:

\[
    \rho^\star=\min_{\beta\in\Delta}\max_{i=0,\ldots,m}
    \sum_{j=0}^m\beta_j \cdot \frac{f(O_i\cup R_j)}{f(O_j\cup R_j)}.
\]

\subsection{The Interval Selection Problem}
\label{sec:rand-interval-selection}

    There is a geometric interpretation of the minimax formulation for $\rho^\star$ in terms of an \emph{interval selection} problem. In this problem, given a distribution over intervals, an interval is selected before the realized interval is drawn. If interval $[\ell_i,r_i]$ is chosen and interval $[\ell_j,r_j]$ is realized, the value is \(
    v(i,j)=\ell_j+\bigl|[\ell_i,r_i]\cap[\ell_j,\infty)\bigr|
\), the sum of the left
endpoint of $[\ell_j,r_j]$ and the length of
$[\ell_i,r_i]$ that extends to the right of $\ell_j$.

\begin{definition}
Let $[\ell_0,r_0],\ldots,[\ell_N,r_N]$ be intervals with
$0\leq\ell_i\leq r_i$, and let $p=(p_0,\ldots,p_N)$ be a probability
distribution over their indices. The \emph{adaptivity gap} of the interval-selection problem is\footnote{We set $\Gamma(\ell,r,p)=1$ when the denominator is zero.} 
\[
\Gamma(\ell,r,p)=
\frac{\displaystyle\max_{i=0,\ldots,N}
      \mathop{\mathbb{E}}\limits_{j\sim p}\!\left[v(i,j)\right]}
     {\displaystyle\mathop{\mathbb{E}}\limits_{j\sim p}\!
      \left[\max_{i=0,\ldots,N}v(i,j)\right]}.
\]
\end{definition}

The numerator is the best expected value of a fixed interval; the denominator
allows the choice to depend on the realized interval. In the reduction from clairvoyant addition-robustness, each interval
corresponds to a prefix of the future elements; its left endpoint $\ell_i=f(R_i)$ is the value
of the prefix alone, while its right endpoint $r_i=f(O_i\cup R_i)$
is the addition-robustness benchmark at prefix $R_i$. Selecting interval $i$
corresponds to committing to $O_i$, and realization $j$ corresponds
to evaluating this choice against prefix $R_j$. The interval value
$v(i,j)$ then lower-bounds $f(O_i\cup R_j)$, the value of this
set of present elements together with the realized prefix. 

\begin{figure}[t]
\centering
\definecolor{cOne}{RGB}{31,119,180}   
\definecolor{cA}{RGB}{255,127,14}     
\definecolor{cB}{RGB}{44,160,44}      

\begin{tikzpicture}[x=6cm,y=0.65cm,
    ext/.style={line width=5pt},
    tick/.style={densely dashed,gray!70}]

  \pgfmathsetmacro{\a}{0.5507946448}
  \pgfmathsetmacro{\b}{0.4704719233}
  \pgfmathsetmacro{\A}{1+\a}
  \pgfmathsetmacro{\B}{1+\a+\b}

  \draw[->,thick] (-0.03,0) -- (\B+0.12,0);
  \foreach \x/\lab in {0/0, 1/1, \A/1+u_1, \B/1+u_1+u_2}{
    \draw (\x,0.08) -- (\x,-0.08) node[below] {$\lab$};
    \draw[tick] (\x,0.1) -- (\x,3.4);
  }

  \draw[ext,cOne] (0,3)  -- (1,3);
  \draw[ext,cA]   (1,2)  -- (\A,2);
  \draw[ext,cB]   (\A,1) -- (\B,1);

  \node[above=1pt] at (0.5,3)        {$I_0$};
  \node[above=1pt] at ({(1+\A)/2},2) {$I_1$};
  \node[above=1pt] at ({(\A+\B)/2},1) {$I_2$};
\end{tikzpicture}
\caption{Interval representation of the hard coverage instance in Figure~\ref{fig:randomized-hard-instance}, with values normalized by $k$.
The future prefixes $\varnothing$, $\{U_0\}$, and $\{U_0,U_1\}$ give the intervals $I_0=[0,1]$, $I_1=[1,1+u_1]$, and $I_2=[1+u_1,1+u_1+u_2]$, respectively.
Their lengths $1,u_1,u_2$ are the normalized gains from selecting all $k$ singleton sets in the first, second, and third groups, respectively.}
\label{fig:rand-hard-intervals}
\end{figure}
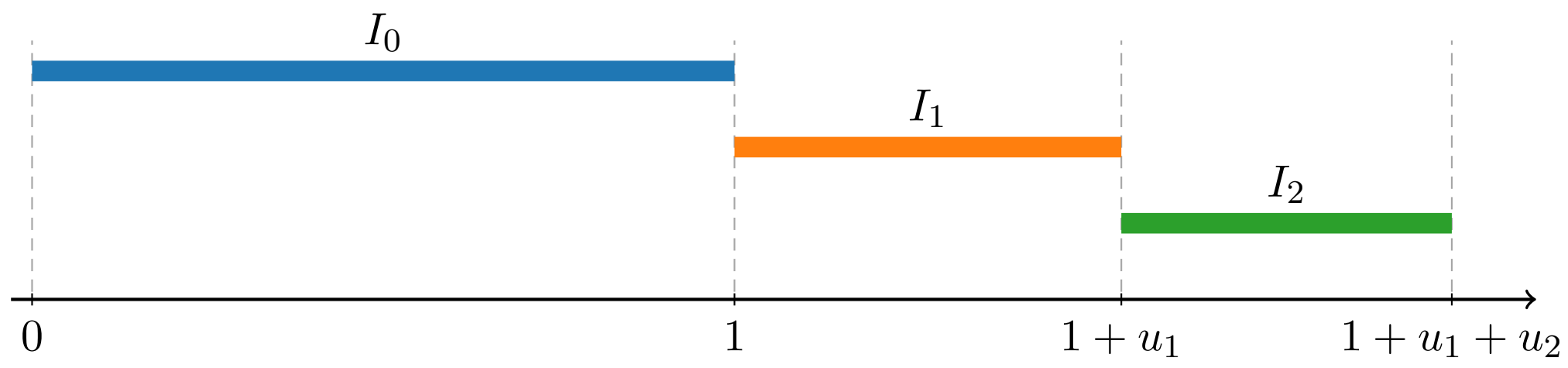

The interval distribution $p$ is obtained from the minimax distribution
$\beta$ by reweighting each probability $\beta_j$ by $1/r_j$ and normalizing.
An adaptivity-gap bound that holds for every distribution $p$ can
therefore be converted, through minimax duality, into one randomized
choice of a set of present elements that performs well against every future
prefix simultaneously. The following lemma makes this connection precise.

\begin{restatable}{lemma}{lossless}
\label{lem:rand-interval-reduction}
Let $\alpha\in[0,1]$. If $\Gamma(\ell,r,p)\geq\alpha$ for every
interval-selection instance, then every instance of clairvoyant
addition-robust  submodular maximization admits an
$\alpha$-addition-robust distribution.
\end{restatable}

It remains to bound the adaptivity gap of the interval selection problem.

\subsection{The Hard Structure for Interval Selection}
\label{sec:rand-hard-structure}

We establish the adaptivity gap of the interval selection problem  through a lossless reduction to the  family of nonincreasing consecutive intervals with the equalizing distribution, for which we establish an adaptivity gap of $\xi^\star$.

\begin{definition} An instance $(\ell,r,p)$ has \emph{nonincreasing consecutive intervals} if
$\ell_0=0$, $\ell_{i+1}=r_i$ for $i<N$, and $r_0-\ell_0\geq\cdots\geq r_N-\ell_N>0$.
Its distribution $p$ is \emph{equalizing} if every fixed interval $i$ has
the same expected value $\mathbb{E}_{j\sim p}[v(i,j)]$.
\end{definition}

Appendix~\ref{sec:rand-interval-proof} derives the unique equalizing
distribution for these intervals. We show that restricting our attention
to this structured family is without loss.

\begin{restatable}{lemma}{chainreduction}
\label{lem:rand-chain-reduction}
Let $\alpha\in[0,1]$. If $\Gamma(\ell,r,p)\geq\alpha$ for every
instance of nonincreasing consecutive intervals with the equalizing
distribution, then $\Gamma(\ell,r,p)\geq\alpha$ for every
interval-selection instance.
\end{restatable}

The final step of the proof of our positive result for randomized algorithms establishes that the adaptivity gap of this family of instances  is $\xi^\star$.

\begin{restatable}{lemma}{sharpadaptive}
\label{lem:rand-chain}
For every instance $(\ell,r,p)$ of nonincreasing consecutive intervals with the equalizing distribution, $\Gamma(\ell,r,p)\geq\xi^\star$.
Equality holds for the three-interval instance $[\ell_0,r_0]=[0,1],$ $[\ell_1,r_1]=[1,1+u_1],$
 and $[\ell_2,r_2]=[1+u_1,1+u_1+u_2]$ with the equalizing distribution, where $u_1$ is the unique root in $(0,1)$ of $\Phi(x)=x^4+x^3-2x^2-3x+2$ and $u_2=u_1^2(1+u_1)$.
\end{restatable}

The equality case in \Cref{lem:rand-chain} is the interval representation of
the hard coverage instance from \Cref{thm:rand_ub_overview}, which is illustrated in Figure~\ref{fig:rand-hard-intervals}. Combining this lemma
with the reductions in Lemmas~\ref{thm:meta}, \ref{lem:rand-interval-reduction},
and~\ref{lem:rand-chain-reduction} proves \Cref{thm:positive_randomized}.

\section{The Tight Deterministic Bound}
\label{sec:sharp-deterministic}

We first give the rows-to-columns obstruction that limits deterministic
algorithms to $\nicefrac34$. We then introduce clairvoyant certificates,
which let us match this bound up to an arbitrarily small loss. The
supporting proofs appear in Appendix~\ref{app:sharp-deterministic}.

\subsection{Impossibility for Deterministic Algorithms}
\label{sec:det-impossibility}

\begin{restatable}{theorem}{impossibilitydeterministic}
\label{thm:det_ub}
    For $\nicefrac{1}{k} \leq \eps \leq 1$, no deterministic $\eps k$-consistent clairvoyant algorithm can achieve an approximation ratio better than $\nicefrac{3}{4}+O(\eps)$ for Maximum Coverage subject to a uniform cardinality constraint $k$.
\end{restatable}

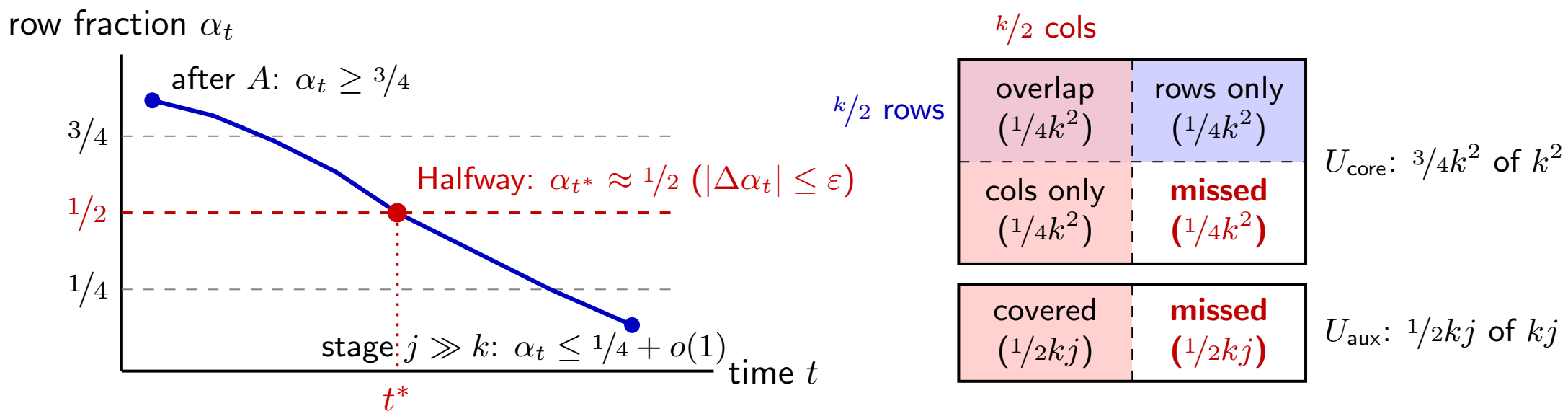
\begin{figure}[t]
\centering
\vspace{-6pt}
\resizebox{0.88\textwidth}{!}{%
\begin{tikzpicture}[font=\sffamily\small]
    \begin{scope}[shift={(0,0.1)}]
        \draw[thick] (0,0) -- (5.8,0) node[right] {time $t$};
        \draw[thick] (0,0) -- (0,3.1) node[above] {row fraction $\alpha_t$};

        \draw[dashed, gray] (0,2.3) node[left, black] {$\nicefrac{3}{4}$} -- (5.4,2.3);
        \draw[thick, dashed, red!75!black] (0,1.55) node[left] {$\nicefrac{1}{2}$} -- (5.4,1.55);
        \draw[dashed, gray] (0,0.8) node[left, black] {$\nicefrac{1}{4}$} -- (5.4,0.8);

        \draw[very thick, blue!75!black]
            (0.3,2.65) -- (0.9,2.5) -- (1.5,2.25) -- (2.1,1.95)
            -- (2.7,1.55) -- (3.4,1.2) -- (4.2,0.8) -- (5.0,0.45);

        \filldraw[blue!75!black] (0.3,2.65) circle (2pt);
        \node[above right, font=\sffamily\footnotesize] at (0.35,2.55) {after $A$: $\alpha_t \ge \nicefrac{3}{4}$};

        \filldraw[red!80!black] (2.7,1.55) circle (2.5pt);
        \draw[dotted, thick, red!80!black] (2.7,1.55) -- (2.7,0) node[below] {$t^*$};
        \node[above right, text=red!80!black, font=\sffamily\footnotesize] at (2.75,1.57) {Halfway: $\alpha_{t^*} \approx \nicefrac{1}{2}$ ($|\Delta \alpha_t| \le \eps$)};

        \filldraw[blue!75!black] (5.0,0.45) circle (2pt);
        \node[below, font=\sffamily\footnotesize] at (3.95,0.52) {stage $j \gg k$: $\alpha_t \le \nicefrac{1}{4}+o(1)$};
    \end{scope}

    \begin{scope}[shift={(8.2,0)}]
        \fill[blue!18] (0,2.15) rectangle (3.4,3.15);
        \fill[red!18] (0,1.15) rectangle (1.7,2.15);
        \fill[purple!22] (0,2.15) rectangle (1.7,3.15);
        \draw[thick] (0,1.15) rectangle (3.4,3.15);
        \draw[dashed] (1.7,1.15) -- (1.7,3.15);
        \draw[dashed] (0,2.15) -- (3.4,2.15);

        \fill[red!18] (0,0) rectangle (1.7,0.95);
        \draw[thick] (0,0) rectangle (3.4,0.95);
        \draw[dashed] (1.7,0) -- (1.7,0.95);

        \node[font=\sffamily\footnotesize, align=center] at (0.85,2.65) {overlap\\($\nicefrac{1}{4}k^2$)};
        \node[font=\sffamily\footnotesize, align=center] at (2.55,2.65) {rows only\\($\nicefrac{1}{4}k^2$)};
        \node[font=\sffamily\footnotesize, align=center] at (0.85,1.65) {cols only\\($\nicefrac{1}{4}k^2$)};
        \node[font=\sffamily\footnotesize\bfseries, text=red!75!black, align=center] at (2.55,1.65) {missed\\($\nicefrac{1}{4}k^2$)};

        \node[font=\sffamily\footnotesize, align=center] at (0.85,0.475) {covered\\($\nicefrac{1}{2}kj$)};
        \node[font=\sffamily\footnotesize\bfseries, text=red!75!black, align=center] at (2.55,0.475) {missed\\($\nicefrac{1}{2}kj$)};

        \node[left, font=\sffamily\footnotesize, text=blue!75!black] at (0,2.65) {$\nicefrac{k}{2}$ rows};
        \node[above, font=\sffamily\footnotesize, text=red!75!black] at (0.85,3.15) {$\nicefrac{k}{2}$ cols};
        \node[right, font=\sffamily\footnotesize] at (3.45,2.15) {$\Ucore$: $\nicefrac{3}{4}k^2$ of $k^2$};
        \node[right, font=\sffamily\footnotesize] at (3.45,0.475) {$\Uaux$: $\nicefrac{1}{2}kj$ of $kj$};
    \end{scope}
\end{tikzpicture}%
}

\vspace{-6pt}
\caption{Left: Consistency forces the row fraction $\alpha_t$ to pass within $\eps$ of $\nicefrac{1}{2}$ at some step $t^*$. Right: The idealized halfway state, with $\nicefrac{k}{2}$ rows and $\nicefrac{k}{2}$ columns, covers $\nicefrac{3}{4}$ of $\Ucore$ and $\nicefrac{1}{2}$ of $\Uaux$, while the complete stage $B_j$ covers both completely.}

\label{fig:det_intuition}
\end{figure}

\begin{proof}[Proof sketch]
    We suppress $O(\eps+1/k)$ losses in this sketch;
    Appendix~\ref{app:det-impossibility} accounts for them explicitly.
    Consider a $k \times k$ grid $\Ucore$ ($k^2$ elements) and an auxiliary pool $\Uaux$ (\Cref{fig:det_intuition}). First, $k$ disjoint \emph{row sets} $A = \{a_1, \dots, a_k\}$ arrive, covering $\Ucore$ row by row. Next, \emph{column sets} arrive in 
    stages $B_1, \dots, B_{\ell}$ for some $\ell \gg k$: 
    in stage $j$, the $k$ disjoint sets in $B_j$ cover $\Ucore$ column by column, and each column $b_{i,j}$ also covers $j$ auxiliary elements in $\Uaux$. 
    Therefore, each column $b_{i,j}$ covers $k+j$ elements, and the collection $B_j$ of $k$ columns in stage $j$ covers $k^2 + kj$ elements.

    Consider any deterministic $\nicefrac{3}{4}$-approximation algorithm $\mathcal{A}$. Let $\alpha_t \in [0,1]$ be the fraction of sets in $A$ that $\mathcal{A}$ selects at time $t$ (\Cref{fig:det_intuition}, left). Right after $A$ arrives, $\OPT$ covers $\Ucore$ ($k^2$ elements), so $\mathcal{A}$ starts with $\alpha_t \ge \nicefrac{3}{4}$. At the end of a stage $j \gg k$, the auxiliary bonus $j$ dominates the core size $k$, so $\OPT = B_j$ covers $k^2 + kj \approx kj$ elements and forces $\ALG$ to switch mostly to columns ($\alpha_t \le \nicefrac{1}{4}+O(k/j)$). Since $\ALG$ is deterministic and changes at most $\eps k$ sets per step, $\alpha_t$ changes by at most $\eps$ per step and must cross a \emph{halfway state} $\alpha_{t^*} \approx \nicefrac{1}{2}$ during some stage $B_j$.

    At step $t^*$, we may upper-bound the value of $\ALG$ by granting it $\nicefrac{k}{2}$ rows from $A$ and $\nicefrac{k}{2}$ distinct columns from $B_j$, up to the suppressed losses. The preceding completed stage provides a benchmark of value at least $(1-O(1/k))f(B_j)$ (\Cref{fig:det_intuition}, right). On $\Ucore$, the $\nicefrac{k}{2}$ rows and $\nicefrac{k}{2}$ columns overlap on one $\nicefrac{k}{2} \times \nicefrac{k}{2}$ quadrant and miss the opposite quadrant, covering $\nicefrac{3}{4}k^2$ of $k^2$ elements. On $\Uaux$, only the $\nicefrac{k}{2}$ selected columns contribute, covering $\nicefrac{1}{2}kj$ of $kj$ elements. Thus the halfway state gets $\nicefrac{3}{4}$ of the complete-stage benchmark on $\Ucore$ and $\nicefrac{1}{2}$ on $\Uaux$:
    \[
        \frac{f(\ALG_{t^*})}{f(\OPT_{t^*})}
        \;\approx\; \frac{\frac{3}{4}k^2 + \frac{1}{2}kj}{k^2 + kj}
        \;\le\; \frac{3}{4}. \qedhere
    \]
\end{proof}

\subsection{Framework for Deterministic Algorithms: Clairvoyant Certificates}
\label{sec:det-framework}

An advantage of a randomized algorithm over a deterministic one is that it can ``hide'' the transition between two solutions by performing it at a random time, so that any fixed time step is affected only with small probability. This is the approach taken in \Cref{sec:sharp-randomized}. In contrast, for a deterministic algorithm, every intermediate solution is compared with the current optimum. Indeed, in the instance used in \Cref{thm:det_ub} and illustrated in \Cref{fig:det_intuition}, a consistent algorithm has to switch gradually from rows to columns, and halfway through the switch its solution covers only $\nicefrac{3}{4}$ of the grid (which is strictly less than what is achievable by a randomized algorithm). Our main result in this part is that this simple example is essentially the only obstacle: a deterministic algorithm can always maintain a $\nicefrac{3}{4}$-approximation while being $O(1)$-consistent, up to an arbitrarily small loss $\eps$.

\begin{theorem}[Deterministic guarantee]
    \label{thm:positive_deterministic}
    For every $\eps\in(0,1)$, there is a deterministic clairvoyant algorithm for monotone submodular maximization under a cardinality constraint $k$ that is $O(\eps^{-3}\log(\nicefrac{1}{\eps}))$-consistent and satisfies
    \[
        f(\ALG_t)\geq(\nicefrac{3}{4}-\eps)f(\OPT_t)
        \qquad\text{at every time }t.
    \]
\end{theorem}

Like its randomized counterpart (\Cref{thm:positive_randomized}), the result is information-theoretic: the algorithm computes exact optima and finds the solution via exhaustive search. The heart of the proof, and the main technical contribution of this section, is a reduction theorem (\Cref{thm:det_reduction}). It shows that a deterministic consistent algorithm exists as soon as a finite object, a \emph{clairvoyant certificate}, exists for every instance of a problem on a single window of the stream, a problem in which consistency plays no role. What remains is to construct certificates: \Cref{thm:det_sqrt2} gives a guarantee of $\nicefrac{1}{\sqrt{2}}$ by a short argument, and \Cref{thm:det_threequarters} gives the tight guarantee $\nicefrac{3}{4}$, with a Lean proof available on GitHub~\citep{BalkanskiDFLNSZ26}. We first introduce the two notions needed to state the reduction: window instances and mixtures.

\paragraph{Window instances.}
The algorithm processes the stream in windows of $O(\eps k)$ consecutive arrivals. Between transitions, it holds a fixed \emph{core} $S$ of about $\kappa=(1-O(\eps))k$ elements that were present at the start of the window, and it outputs $S$ together with a buffer of recent arrivals. This buffer includes the arrivals of the current window and retains those needed from the preceding window until the transition finishes. Fix a window, let $V_{\mathrm{now}}$ be the set of elements present at its start, $V_{\mathrm{future}}=(v_1,\dots,v_m)$ its arrivals in order, and $R_j=\{v_1,\dots,v_j\}$, with $R_0=\varnothing$. The best core in hindsight after $j$ arrivals defines the benchmark at \emph{stage} $j$,
\begin{equation}
    \label{eq:det-benchmark}
    f(O_j\cup R_j),
    \qquad\text{where}\qquad
    O_j\in\operatorname*{arg\,max}_{O\subseteq V_{\mathrm{now}},\,|O|\le\kappa} f(O\cup R_j).
\end{equation}
It is at least $(1-O(\eps))f(\OPT_t)$ at the time $t$ of the $j$-th arrival, so a core that is $\alpha$-competitive against it is an $(\alpha-O(\eps))$-approximation: dropping $k-\kappa$ elements from $\OPT_t$ loses at most a $(k-\kappa)/k$ fraction of its value, and what remains inside $V_{\mathrm{now}}$ is a feasible core. The tuple $(f,V_{\mathrm{now}},V_{\mathrm{future}},\kappa)$, together with a choice of the sets $O_0,\dots,O_m$, which we call the \emph{candidates}, is a \emph{window instance} with stages $0,\dots,m$; it is exactly an instance of the clairvoyant addition-robust problem that \Cref{sec:rand-robustness} studies for randomized algorithms.  The candidates are known in advance, but a consistent algorithm cannot jump between them; the question is which core to hold.

\paragraph{Mixtures.}
For randomized algorithms, a distribution over the candidates answers this question (this is the clairvoyant addition-robustness of \Cref{sec:rand-robustness}): outputting $O_i$ with probability $p_i$ yields expected value $\sum_i p_i f(O_i\cup R_j)$ at stage $j$, and by the reduction of \Cref{thm:meta} such a distribution is all that is needed. In contrast, a deterministic algorithm cannot sample. It has to hold a single set that blends the candidates, and since $f$ is submodular, the value of a blend is governed by the ``overlaps'' of the candidates, not by the average of their values. It is instructive to see what happens on the grid of \Cref{thm:det_ub}, which the rows and the columns each cover completely. Outputting the rows with probability $\nicefrac{1}{2}$ or the columns with probability $\nicefrac{1}{2}$ covers the whole grid either way. If instead every row and every column is picked independently with probability $\nicefrac{1}{2}$, each cell is covered with probability $\nicefrac{3}{4}$, so only $\nicefrac{3}{4}$ of the grid is covered in expectation; this is the value of the multilinear extension at the fractional point that puts weight $\nicefrac{1}{2}$ on each of the two candidates. We will work with the simpler quantity~\eqref{eq:H-function} below, which gives a lower bound on the multilinear extension: it draws the row candidate and the column candidate independently, each with probability $\nicefrac{1}{2}$, and merges them; with probability $\nicefrac{1}{4}$ nothing is drawn, so this again covers $\nicefrac{3}{4}$ of the grid in expectation.

The clairvoyant-certificate framework handles this difficulty in a powerful and modular way. It charges for the overlaps once, in the value~\eqref{eq:H-function} it assigns to a blend, and \Cref{thm:det_reduction} then turns a randomized-looking guarantee into a deterministic consistent algorithm with the same guarantee. As in the randomized case, designing the algorithm thereby amounts to proving expectation bounds on one window; the rounding, the timing of the transitions, and the accounting for consistency are handled by the general reduction theorem.

Concretely, a \emph{mixture} is a probability vector $p\in\Delta$ over the candidates, where $\Delta$ is the simplex of distributions on $\{0,\dots,m\}$, and its value at stage $j$ is
\begin{equation}
    H_j(p)=\mathbb{E}\Big[f\Big(R_j\cup\bigcup\nolimits_{i\in I(p)}O_i\Big)\Big],
    \label{eq:H-function}
\end{equation}
where $I(p)$ contains each index $i$ independently with probability $p_i$. By submodularity, $H_j(p)$ is at most the randomized value $\sum_i p_i \cdot f(R_j\cup O_i)$; Appendix~\ref{d2:sec:local} proves this comparison and the lower bound on the multilinear extension. On the other hand, one set of about $\kappa$ elements can simultaneously attain these guarantees at every stage, up to an $\eps$ loss; this is the rounding step of the reduction. An advantage of the above definition over the multilinear extension is that it is much less sensitive to small changes in the probabilities. In particular, changing the probability \(p_i\) assigned to \(O_i\) by \(\epsilon\) changes the function value by at most \(\epsilon f(O_i)\). For the multilinear extension, in contrast, the effect can be much larger. To see this, consider the submodular function \(f\) that satisfies \(f(S)=1\) for every nonempty set \(S\) and \(f(\emptyset)=0\). If \(O_i\) has cardinality \(k\), then increasing \(p_i\) from \(0\) to \(\epsilon\) increases the multilinear-extension value from \(0\) to
\[
1-(1-\epsilon)^k,
\]
which can be arbitrarily close to \(1\) even when \(\epsilon\) is a small constant.

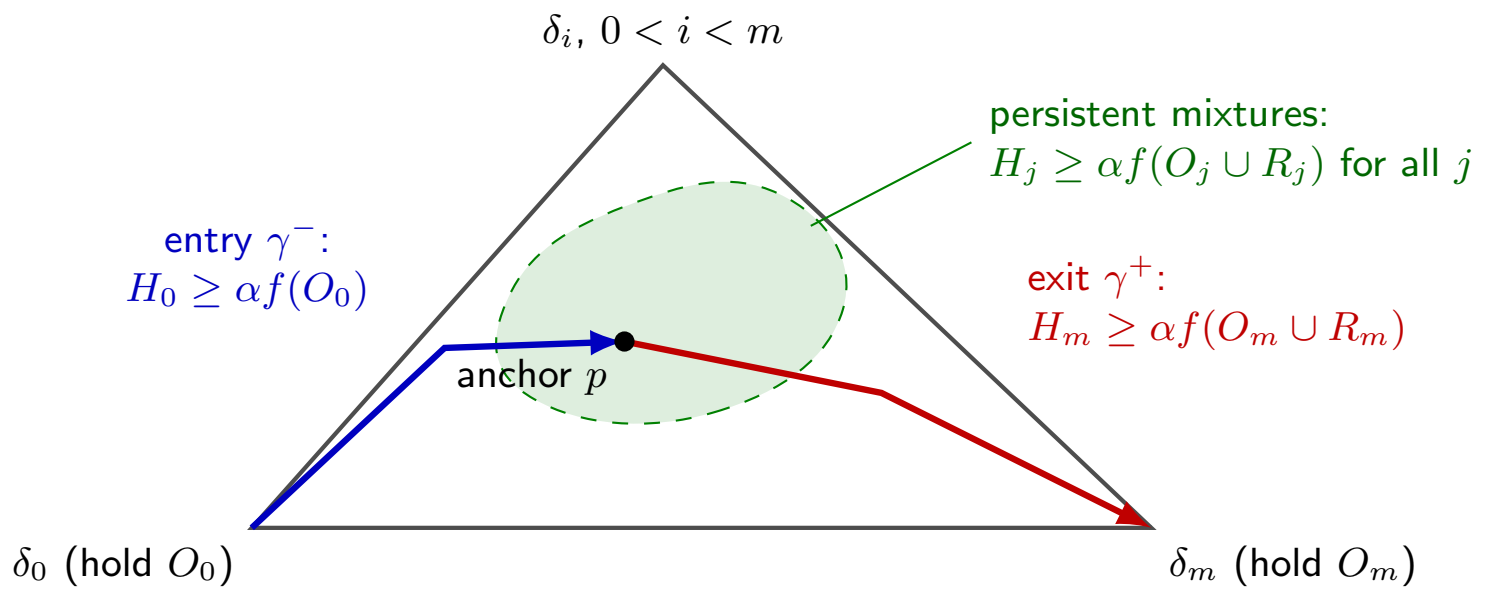
\begin{figure}[t]
    \centering
    \begin{tikzpicture}[font=\sffamily\small, scale=0.85, every node/.style={transform shape}]
        \coordinate (O0) at (0,0);
        \coordinate (Om) at (7,0);
        \coordinate (Oi) at (3.2,3.6);
        \draw[thick, black!70] (O0) -- (Om) -- (Oi) -- cycle;
        \node[below left, font=\sffamily\footnotesize] at (O0) {$\delta_0$ (hold $O_0$)};
        \node[below right, font=\sffamily\footnotesize] at (Om) {$\delta_m$ (hold $O_m$)};
        \node[above, font=\sffamily\footnotesize] at (Oi) {$\delta_i$, $0<i<m$};
        \fill[green!50!black, opacity=0.13] plot[smooth cycle, tension=0.8] coordinates {(2.2,1.0) (3.6,0.9) (4.6,1.7) (4.1,2.6) (2.9,2.5) (2.0,1.8)};
        \draw[green!50!black, dashed] plot[smooth cycle, tension=0.8] coordinates {(2.2,1.0) (3.6,0.9) (4.6,1.7) (4.1,2.6) (2.9,2.5) (2.0,1.8)};
        \draw[green!50!black, thin] (4.35,2.35) -- (5.6,3.0);
        \node[green!40!black, align=left, font=\sffamily\footnotesize, anchor=west] at (5.6,3.0) {persistent mixtures:\\$H_j\ge\alpha f(O_j\cup R_j)$ for all $j$};
        \coordinate (P) at (2.9,1.45);
        \draw[very thick, blue!75!black, -latex] (O0) -- (1.5,1.4) -- (P);
        \node[blue!75!black, align=center, font=\sffamily\footnotesize, anchor=east] at (1.05,2.05) {entry $\gamma^-$:\\$H_0\ge\alpha f(O_0)$};
        \draw[very thick, red!75!black, -latex] (P) -- (4.9,1.05) -- (Om);
        \node[red!75!black, align=left, font=\sffamily\footnotesize, anchor=west] at (5.9,1.75) {exit $\gamma^+$:\\$H_m\ge\alpha f(O_m\cup R_m)$};
        \filldraw[black] (P) circle (2pt) node[below left, font=\sffamily\footnotesize] {anchor $p$};
    \end{tikzpicture}
    \caption{A clairvoyant certificate (\Cref{def:det_certificate}) in the simplex $\Delta$ of mixtures; the top vertex stands for the intermediate candidates. The anchor $p$ is persistent, the entry path from $\delta_0$ is safe at stage $0$, the exit path to $\delta_m$ is safe at stage $m$, and the number of segments governs the consistency.}
    \label{fig:det_certificate}
\end{figure}

\paragraph{Clairvoyant certificates.}
A mixture that is $\alpha$-competitive against the benchmark \eqref{eq:det-benchmark} at every stage is not enough on its own. The algorithm reaches $O_0$ at the start of the window and then moves into the mixture. At the end of the window, it has to move on to $O_m$, from which the reduction connects to the next window. The intermediate mixtures along both moves have to be good as well. Each move is judged against one benchmark: the entry, made during the first few arrivals of the window, is evaluated against the benchmark at stage $0$, without credit for those arrivals; the exit, made after the last arrival, is evaluated against the benchmark at stage $m$. One benchmark per move suffices because the reduction moves only while the optimum is nearly flat, or while the recent arrivals alone are a $\nicefrac{3}{4}$-approximation. Writing $\delta_i\in\Delta$ for the vertex that puts all weight on $O_i$, we arrive at the following notion, illustrated in \Cref{fig:det_certificate}.

\begin{definition}[Clairvoyant certificate]
    \label{def:det_certificate}
    An $(\alpha,L)$-\emph{certificate} for a window instance consists of a mixture $p\in\Delta$, the \emph{anchor}, and two continuous piecewise-affine paths $\gamma^-,\gamma^+\colon[0,1]\to\Delta$ with $\gamma^-(0)=\delta_0$, $\gamma^-(1)=p=\gamma^+(0)$, and $\gamma^+(1)=\delta_m$, such that
    \begin{align*}
        H_j(p)           & \ge\alpha\, f(O_j\cup R_j) &  & \text{for all } j=0,\dots,m, &  & \text{(persistence)} \\
        H_0(\gamma^-(s)) & \ge\alpha\, f(O_0)         &  & \text{for all } s\in[0,1],   &  & \text{(safe entry)}  \\
        H_m(\gamma^+(s)) & \ge\alpha\, f(O_m\cup R_m) &  & \text{for all } s\in[0,1],   &  & \text{(safe exit)}
    \end{align*}
    and such that the two paths consist of at most $L$ segments in total.
\end{definition}

\paragraph{The reduction theorem.}
We can now state the main technical result of this section.

\begin{theorem}[Reduction from consistency to clairvoyant certificates]
    \label{thm:det_reduction}
    Let $0<\alpha\le\nicefrac{3}{4}$ and suppose that every window instance with stages $0,\dots,m$ admits an $(\alpha,L)$-certificate with $L\le C\,(m+1)$ for a constant $C$. Then, for every $\eps\in(0,1)$, there is a deterministic clairvoyant $O(\eps^{-3}\log(\nicefrac{1}{\eps}))$-consistent algorithm that satisfies
    \[
        f(\ALG_t)\geq(\alpha-\eps)f(\OPT_t)
        \qquad\text{at every time }t,
    \]
    where the constant in the consistency bound depends only on $C$.
\end{theorem}

The proof, given in Appendix~\ref{d2:sec:transfer}, has two parts: a schedule, which decides when the transitions take place, and a rounding procedure, the main novelty, which turns a certificate into an integral anchor and two lists of operations.  We describe the schedule first, then explain the rounding that supplies its transition lists.

\begin{figure}[t]
    \centering
    \begin{tikzpicture}[font=\sffamily\footnotesize, x=1cm, y=1cm]
        \node[anchor=east] at (-0.15,3.6) {blocks};
        \foreach \x/\lab in {0/{$i-1$}, 4.8/{$i$}, 9.6/{$i+1$}} {
        \draw[thick, black!70, fill=black!5] (\x,3.3) rectangle (\x+4.8,3.9);
        \node at (\x+2.4,3.6) {block \lab: $\Theta(\eps k)$ arrivals};
        }
        \foreach \x in {1.2,2.4,...,13.2} \draw[black!60] (\x,3.3) -- (\x,3.12);
        \node[anchor=east, align=right] at (-0.15,2.7) {transition\\interval};
        \foreach \x in {4.8,9.6,10.8} \draw[fill=orange!30, draw=orange!80!black] (\x,2.45) rectangle (\x+1.2,2.95);
        \foreach \x in {1.2,6.0,12.0} \draw[fill=red!20, draw=red!75!black, thick] (\x,2.45) rectangle (\x+1.2,2.95);
        \node[text=orange!80!black, font=\sffamily\scriptsize, anchor=north] at (10.8,2.42) {optimum grows by more than $1+\eps$};
        \node[anchor=east] at (-0.15,1.45) {core};
        \draw[very thick, blue!75!black] (0,1.45) -- (1.2,1.45);
        \draw[very thick, blue!75!black] (2.4,1.45) -- (6.0,1.45) node[midway, above=2pt, text=blue!75!black] {hold anchor $i-1$};
        \draw[very thick, blue!75!black] (7.2,1.45) -- (12.0,1.45) node[midway, above=2pt, text=blue!75!black] {hold anchor $i$};
        \draw[very thick, blue!75!black] (13.2,1.45) -- (14.4,1.45);
        \foreach \x in {1.2,6.0,12.0} \draw[fill=red!20, draw=red!75!black, thick] (\x,1.2) rectangle (\x+1.2,1.7);
        \node[text=red!75!black, font=\sffamily\scriptsize, anchor=north] at (6.6,1.15) {exit list of window $i-1$, segment $O_m\to O_0$, entry list of window $i$};
        \draw[decorate, decoration={brace, mirror, amplitude=4pt}, thick, blue!75!black] (1.2,0.45) -- (6.0,0.45) node[midway, below=5pt, text=blue!75!black] {window $i-1$};
        \draw[decorate, decoration={brace, mirror, amplitude=4pt}, thick, blue!75!black] (6.0,0.45) -- (12.0,0.45) node[midway, below=5pt, text=blue!75!black, align=center] {window $i$: one window instance,\\$V_{\mathrm{now}}$ fixed at its start};
        \draw[decorate, decoration={brace, mirror, amplitude=4pt}, thick, blue!75!black] (12.0,0.45) -- (14.4,0.45) node[midway, below=5pt, text=blue!75!black] {window $i+1$};
    \end{tikzpicture}
    \caption{The schedule of the reduction. Blocks of $\Theta(\eps k)$ arrivals are divided into $\lceil\nicefrac{4}{\eps}\rceil$ subintervals of $\Theta(\eps^2k)$ arrivals. In each block, the transition (red) takes place in the first subinterval during which the optimum grows by at most a factor of $1+\eps$; subintervals in which it grows faster (orange) are skipped. A window runs from one transition to the next; in between, the core holds the rounded anchor of the window, and during a transition, the algorithm performs $O(\nicefrac{L}{\eps^2})$ changes per arrival.}
    \label{fig:det_schedule}
\end{figure}
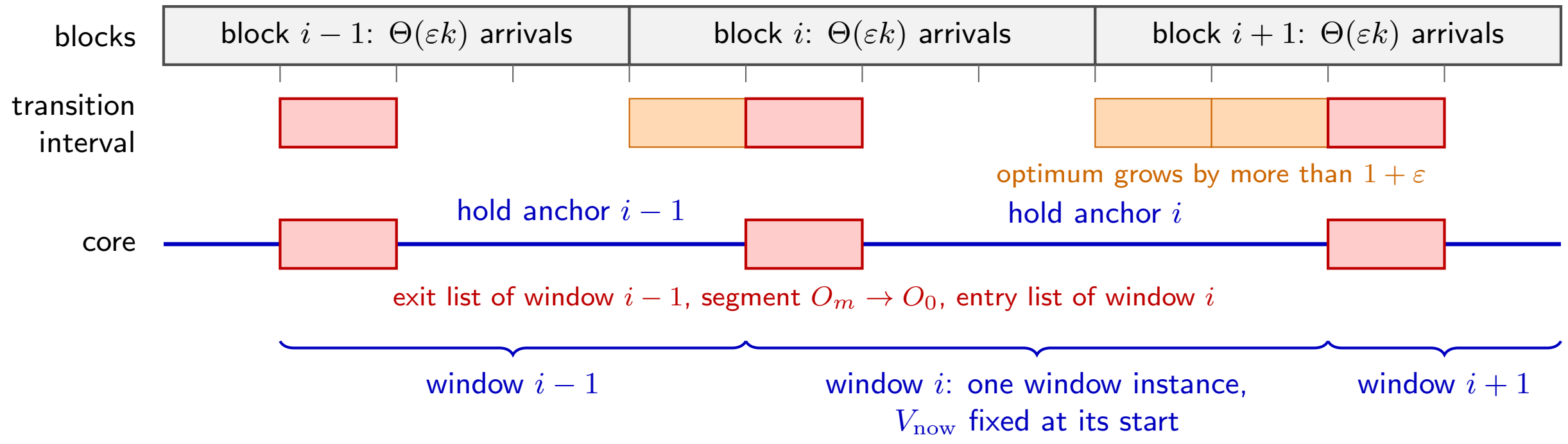

\emph{Scheduling.}
The time structure is shown in \Cref{fig:det_schedule}. The stream is cut into blocks of $\Theta(\eps k)$ arrivals, and each block hosts one transition, placed using foresight in the first subinterval of $\Theta(\eps^2k)$ arrivals during which $f(\OPT_t)$ grows by at most a factor of $1+\eps$; this is why one benchmark per move suffices. If there is no such subinterval, the optimum has at least quadrupled over the block, and the arrivals of the block, which are in the output, are by themselves a $\nicefrac{3}{4}$-approximation. We therefore choose the last subinterval of the block for the transition. A window runs from one transition to the next; in between, the core holds the rounded anchor and persistence provides the approximation. A transition executes the exit list of the old window, a connecting segment from the old $O_m$ to the new $O_0$, and the entry list of the new window; the connecting segment retains at least a $\nicefrac{3}{4}$ fraction of the common benchmark met by its endpoints. Only $O(\nicefrac{1}{\eps})$ stages of a window need to be checked, one for each increase of $f(R_j)$ by $\eps f(O_0)$, so $L=O(\nicefrac{1}{\eps})$ and the $O(L\kappa)$ operations of a transition cost $O(\eps^{-3})$ changes per arrival; the logarithm in the consistency bound of \Cref{thm:det_reduction} covers very small $k$, for which one simply recomputes an optimum after every arrival. The factor $\nicefrac{3}{4}$ enters only through the connecting segment and the fallbacks to the arrivals, which is why the theorem is stated for $\alpha\le\nicefrac{3}{4}$.

\emph{Rounding.}
A certificate gives expectation guarantees at the anchor for every stage and at every point of two continuous paths, whereas the algorithm needs one integral core and two lists of insertions and deletions, every prefix of which is good. The first step is a discretization: the anchor is tested at the $O(\nicefrac{1}{\eps})$ retained stages, and each of the $L$ segments is sampled at $O(\nicefrac{1}{\eps})$ mesh points, between which $H$ changes by at most $\eps f(O_0)$. This leaves $M=O(\nicefrac{L+1}{\eps})=O(\eps^{-2})$ tests, a number that depends only on $\eps$, not on $k$ or the ground set; this is important for the upcoming union bound. The second step uses common thresholds for rounding. Let $z(p)=\sum_i p_i\mathbf{1}_{O_i}$ be the fractional vector within the budget $\kappa$. Draw one uniform threshold $U_e$ per element once and for all, and round every mixture $p$ to the set of elements with $U_e\le z_e(p)$, so that the anchor and all mesh points are rounded with the same randomness. At each test, the rounded set is worth at least $H_j(p)$ in expectation, since its expected value is the multilinear extension at $z(p)$. After the few elements of large marginal value have been forced in, a concentration bound for submodular functions makes the test fail with probability well below $\nicefrac{1}{M}$ provided $k$ exceeds a sufficiently large polynomial in $\nicefrac{1}{\eps}$; the sizes of all sampled sets are controlled in the same way. The expected number of threshold crossings is $O(L\kappa)$, and Markov's inequality bounds the probability of exceeding a constant multiple of this number. Together with the union bound over the tests, this leaves a realization that satisfies every value, size, and operation bound; we fix it, and the anchor and both lists become deterministic. The third step is where the common thresholds pay off. Along a segment, every $z_e$ is affine, so an element crosses its threshold at most once per segment. The resulting lists have $O(L\kappa)$ operations regardless of the mesh. This operation bound is small enough to achieve the desired consistency. Between adjacent mesh points, the two rounded sets both contain the rounding of the coordinatewise smaller type vector, whose value is certified, so performing the deletions before the insertions keeps every intermediate set good, and a list can be paused after any operation.

\paragraph{Constructing certificates.}
It remains to construct certificates, which we do at two levels.

\begin{theorem}[Certificates at $\nicefrac{1}{\sqrt{2}}$]
    \label{thm:det_sqrt2}
    Every window instance with stages $0,\dots,m$ admits a $(\nicefrac{1}{\sqrt{2}},3)$-certificate.
\end{theorem}
Together with \Cref{thm:det_reduction}, this certificate yields a deterministic $(\nicefrac{1}{\sqrt{2}}-\eps)$-approximation; Appendix~\ref{d2:sec:half} proves the certificate theorem. Its anchor is the half-mixture
\(
    p=\tfrac{1}{2}(\delta_0+\delta_c),
\)
where we choose the stage $c$ carefully to balance the value of the mixture before and after that stage. The choice makes the anchor competitive throughout the window.

The two paths complete the certificate by showing how to enter and leave the anchor without a large drop in value. The entry is the straight path from $\delta_0$ to $p$, and the exit goes from $p$ through whichever of $\delta_0$ and $\delta_c$ is more valuable at the final stage, and then to $\delta_m$. This leads to three segments in total and we show that they possess the required guarantees. Thus, together with the anchor they form a certificate with three segments.

One can show that half-mixtures cannot do better than $1/\sqrt{2}$, and the tight bound is more complex and uses anchors whose weights are tuned to the instance. 

\begin{theorem}[Certificates at $\nicefrac{3}{4}$]
    \label{thm:det_threequarters}
    Every window instance with stages $0,\dots,m$ admits a $(\nicefrac{3}{4},m+3)$-certificate.
\end{theorem}

\Cref{thm:positive_deterministic} follows from \Cref{thm:det_reduction,thm:det_threequarters}, since $m+3\le 3(m+1)$. The proof of \Cref{thm:det_threequarters} is a Lean certificate, formally checked by the Lean proof assistant; a clean, human-readable proof remains open. An AI-generated natural-language proof and the Lean certificate are available on GitHub~\citep{BalkanskiDFLNSZ26}.


\section*{AI Disclosure}

We acknowledge the use of several state-of-the-art LLM models (such as ChatGPT Astra, Claude Opus, and Gemini Argon) as follows.

For the information-theoretic bound for randomized algorithms (Thm.~\ref{thm:rand_ub_overview} and Thm.~\ref{thm:positive_randomized}), we seeded the models with our earlier paper \citep{DuttingFLNSZ25} and the adjusted LP formulation in \eqref{lp:randomized}. We spent significant effort in interpreting and refactoring the resulting proof, and making it as insightful as possible. In particular, the interpretation of the resulting optimization problem as an interval selection problem is based on our reading of the AI-generated proofs, as is the formalization of the connection through the adaptivity gap.

To obtain the information-theoretic bounds for deterministic algorithms (Thm.~\ref{thm:det_ub} and Thm.~\ref{thm:positive_deterministic}) we started from a human-generated impossibility of $3/4$ along with a human-generated proof sketch that obtains a $0.57$ approximation (see~\cite{BalkanskiDFLNSZ26prompt} for that sketch). Iterative prompting allowed us to improve the lower bound and eventually close the gap, showing a tight $3/4$ approximation.
The first AI-generated proof of the $3/4$ guarantee was about $100$ pages long and near-impossible to parse and understand. Giving this proof the human-understandable form of Section~\ref{sec:sharp-deterministic} is our main achievement. To reach this point, we used an AI-assisted refinement approach. We first asked the model to generate a simpler proof that beats $1-1/e$, which led to the $1/\sqrt{2}$ proof. We subsequently asked the model to abstract the two proofs in a similar manner as the (clairvoyant) addition-robustness notion abstracts the proof for randomized algorithms. This led to a human-understandable proof using the concept of certificates, which formed the basis of the human-written proof outline of Section~\ref{sec:det-framework}.

We have carefully checked the correctness of the certificates framework as well as the elementary construction that leads to the $1/\sqrt{2}$ certificate (included in Appendix~\ref{d2:sec:deterministic-two}). We could not find any meaningful insights in the improved $3/4$ certificate, and therefore decided to delegate that proof to an external appendix and Lean proof (see \cite{BalkanskiDFLNSZ26}).

\bibliographystyle{plainnat}
\bibliography{references}

\appendix 

\section{Proofs for the Tight Randomized Bound}
\label{app:sharp-randomized}

This appendix proves the results stated in Section~\ref{sec:sharp-randomized}.
We first verify that $\xi^\star$ is well defined in Appendix~\ref{app:unique}
and complete the coverage impossibility proof in
Appendix~\ref{app:rand-lower-bound}, including the finite-$k$ error terms
and the conversion to an unweighted instance.

The algorithmic proofs follow the reductions in Figure~\ref{fig:rand-reductions}.
Appendix~\ref{app:rand-checkpoints} converts clairvoyant addition robustness
into consistency, and Appendix~\ref{app:rand-addition-proofs} derives
addition robustness from an interval-selection bound. We establish that
bound by reducing general intervals to consecutive intervals in
Appendix~\ref{sec:rand-interval-proof} and analyzing the latter in
Appendix~\ref{app:rand-structured}. Finally,
Appendix~\ref{app:rand-combination} combines these ingredients to prove
Theorem~\ref{thm:positive_randomized}.

\subsection{Properties of the Polynomial \texorpdfstring{$\Psi$}{Psi}}
\label{app:unique}

We begin by checking that the constant used in Section~\ref{sec:sharp-randomized}
is well defined.
        \begin{proposition}
            The polynomial $\Psi(\xi)=89\xi^4-44\xi^3-19\xi^2-3\xi+1$ has a unique root $\xi^\star\approx 0.7941970147$ in $(\nicefrac{1}2,1)$.
        \end{proposition}
        \begin{proof}

The polynomial $\Psi$ is continuous and has opposite signs at
$\nicefrac12$ and $1$:
\begin{align*}
\Psi\!\left(\tfrac12\right) &= \tfrac{89}{16}-\tfrac{44}{8}-\tfrac{19}{4}-\tfrac32+1
= -\tfrac{83}{16} < 0,\\
\Psi(1) &= 89-44-19-3+1 = 24 > 0.
\end{align*}
The intermediate value theorem therefore gives a root in $(\nicefrac12,1)$.
To prove uniqueness, we show that $\Psi$ is strictly convex on this interval.
Its first three derivatives are
\begin{align*}
\Psi'(\xi)&=356\xi^3-132\xi^2-38\xi-3,\\
\Psi''(\xi) &= 1068\xi^2-264\xi-38,\\
\Psi'''(\xi) &= 2136\xi-264.
\end{align*}
For $\xi\ge\nicefrac12$, we have $\Psi'''(\xi)\ge 804>0$, so $\Psi''$ is increasing on
$[\nicefrac12,1]$. Since $\Psi''\!\left(\tfrac12\right)=267-132-38=97>0$, it follows that
$\Psi''>0$ on $[\tfrac12,1]$, so $\Psi$ is strictly convex there.
Suppose, for a contradiction, that $\Psi$ has two roots $\xi_1<\xi_2$
in $(\tfrac12,1)$.
Write $\xi_1=\lambda\cdot\nicefrac12+(1-\lambda)\,\xi_2$ with $\lambda\in(0,1)$.
By strict convexity,
\[
\Psi(\xi_1) < \lambda\,\Psi\!\left(\tfrac12\right)+(1-\lambda)\,\Psi(\xi_2)
= \lambda\,\Psi\!\left(\tfrac12\right) < 0,
\]
contradicting $\Psi(\xi_1)=0$. Hence the root is unique.

For the numerical value, direct evaluation at rational endpoints gives
$\Psi(0.7941970146)<0<\Psi(0.7941970148)$, locating the unique root in this interval.
\end{proof}
\subsection{Impossibility for Randomized Algorithms}
\label{app:rand-lower-bound}

We give the full analysis of the construction in
Section~\ref{sec:rand-impossibility}. The limiting linear program in the
proof sketch determines the constant $\xi^\star$; here we account for the
recourse budget and the exact cardinality constraint, then remove the weights.

\hardrand*

\begin{proof}
        For $k=1$, the statement is trivial since $\eps=1$ and no approximation
        ratio exceeds $1$. Fix $k\geq2$ and $\eps\in[1/k,1]$. We first prove
        the bound for a weighted coverage instance.

        Let $u_1$ be the unique root in $(0,1)$ of
        \(
            \Phi(x)=x^4+x^3-2x^2-3x+2,
        \) and set $u_2=u_1^2(1+u_1)$. The parameter analysis in the proof of
        Lemma~\ref{lem:rand-chain} establishes $0<u_2<u_1<1$ independently of
        the construction below. Numerically,
        \[
            u_1\approx 0.550794645,
            \qquad
            u_2\approx 0.470471923.
        \]

        \paragraph{The instance.}
        Partition a universe $U$ into three groups $U_0,U_1,U_2$ of $k$
        elements each, assigning weight $u_i$ to every element of $U_i$,
        where $u_0=1$. The stream consists of the $3k$ singleton subsets of
        $U$, in any order, followed by the large sets $U_0$ and $U_1$.
        This is the instance in Figure~\ref{fig:randomized-hard-instance}.

        Relabel the last three time steps as $0$, $1$, and $2$: immediately before
        $U_0$ arrives, immediately after $U_0$ arrives, and immediately after
        $U_1$ arrives. Write $o_i$ for the optimal value and $a_i$ for the
        algorithm's expected value at these time steps.

        At time $0$, an optimal solution selects all $k$ singletons from
        $U_0$, so $o_0=k$. At time $1$, it selects $U_0$ and $k-1$ singletons
        from $U_1$, giving $o_1=k+u_1(k-1)$. At time $2$, it selects both
        large sets and $k-2$ singletons from $U_2$, giving
        $o_2=k(1+u_1+u_2)-2u_2$.

        Fix a randomized $\eps k$-consistent algorithm, and let $k_i$ be the
        expected number of singletons from $U_i$ that it holds at time $0$.
        Granting it the large sets as they arrive and using consistency to
        bound the additional singleton sets gives
        \begin{align*}
            a_0 &= k_0+u_1 k_1+u_2 k_2,\\
            a_1 &\leq k+u_1(k_1+\eps k)+u_2 k_2,\\
            a_2 &\leq k+u_1 k+u_2(k_2+2\eps k).
        \end{align*}
        The terms involving $\eps$ account for at most $\eps k$ newly selected
        singleton sets per update. At time $1$, each new singleton outside
        $U_0$ has weight at most $u_1$; at time $2$, only singletons in $U_2$,
        each of weight $u_2$, can contribute beyond
        the two large sets.

        \paragraph{Bounding the approximation ratio.}
        Any approximation ratio $\rho$ must satisfy $a_i\geq\rho o_i$ at all
        three time steps. Substituting the bounds above yields a linear-program
        relaxation in the expected counts $k_i$, which need not be integers.
        Normalize them as $\alpha_i=k_i/k$ and replace both $\eps$ and $1/k$
        by $\hat\eps=\eps+1/k$; this only relaxes the constraints. It suffices
        to consider $0\leq\rho\leq1$, so the program is
        \[
            \begin{aligned}
                \text{Maximize}\quad &\rho\\
                \text{subject to}\quad
                &\alpha_0+\alpha_1+\alpha_2\leq 1,\\
                &\alpha_0+u_1\alpha_1+u_2\alpha_2\geq\rho,\\
                &1+u_1(\alpha_1+\hat\eps)+u_2\alpha_2
                    \geq\rho\left(1+u_1-{u_1}\hat \eps\right),\\
                &1+u_1+u_2(\alpha_2+2\hat\eps)
                    \geq\rho\left(1+u_1+u_2-{2\hat \eps u_2}\right),\\
                &\alpha_i\geq 0\quad\text{for }i=0,1,2,\qquad 0\leq\rho\leq1.
            \end{aligned}
        \]
        We bound its optimal value by $\xi^\star+O(\hat\eps)=\xi^\star+O(\eps)$.
        We may assume $\alpha_0+\alpha_1+\alpha_2=1$, since increasing
        $\alpha_0$ to fill unused capacity only relaxes the first
        approximation constraint. Take a convex combination of the three
        approximation constraints with weights
        \[
            \mu_0=u_2,
            \qquad
            \mu_1=\frac{u_2}{u_1}-u_2,
            \qquad
            \mu_2=1-\frac{u_2}{u_1}.
        \]
        The resulting inequality is
        \[
             1+u_1-u_2 + (1+\rho)\,\overline  \eps  \;\geq\; \rho \left(1+u_1+u_2-u_1u_2-\frac{u_2^2}{u_1}\right),
        \]
        where $\overline \eps=\hat \eps u_2\left(3-u_1-\nicefrac{2u_2}{u_1}\right)\approx 0.3485 \hat\eps$.
        Put $D=1+u_1+u_2-u_1u_2-u_2^2/u_1>0$; this is a convex combination of the positive limiting benchmark values. Using $\rho\leq1$ gives the uniform bound
        \[
            \rho \le \frac{1+u_1-u_2}{D}+\frac{2\overline\eps}{D}
            =\frac{1+u_1-u_2}{D}+O(\eps),
        \]
        where we use $1/k\leq\eps$ to obtain $\hat\eps\leq2\eps$.
        By our choice of $u_1$ and $u_2$,
        \begin{align*}
            \left(1+u_1+u_2-u_1u_2-\frac{u_2^2}{u_1} \right)-(1+u_1-u_2)(1+u_1u_2)=u_2\Phi(u_1)=0,
        \end{align*}
        so that $\rho \le \frac{1}{1+u_1u_2} + O(\eps) = \xi^\star+O(\eps)$.
        Indeed, substituting $u_2=u_1^2(1+u_1)$ and using the equation for $u_1$
        shows that $1/(1+u_1u_2)$ lies in $(\nicefrac12,1)$ and satisfies
        $\Psi(\xi)=89\xi^4-44\xi^3-19\xi^2-3\xi+1=0$.

        \paragraph{Removing the weights.}
        To obtain an unweighted coverage instance, choose rational weights
        $0<\tilde u_2<\tilde u_1<1$ with $|\tilde u_i-u_i|<\eps$ for $i=1,2$.
        Such weights exist for every real $\eps>0$ by density of the rationals; they need not satisfy $\tilde u_2=\tilde u_1^2(1+\tilde u_1)$.
        This changes the coverage of any solution, and hence each optimal value,
        by at most $2\eps k$, while all three optimal values remain at least $k$.
        The weighted bound applies to every feasible sequence of solutions, including one chosen for the perturbed instance. Each of the three expected approximation ratios changes by at most $4\eps$, so the bound changes by only $O(\eps)$.
        Multiply all weights by a common denominator and replace each
        weighted universe element by the corresponding number of
        unit-weight copies, with the same incidences to the stream sets.
        This preserves all approximation ratios and the cardinality and
        consistency constraints.
    \end{proof}

    \begin{remark}
        For any fixed consistency bound $C$ and any $\delta>0$,
        choosing $k$ sufficiently large yields an unweighted coverage
        instance on which no randomized $C$-consistent clairvoyant
        algorithm can maintain an approximation ratio better than
        $\xi^\star+\delta$. Thus the limiting impossibility constant
        matches the guarantee in Theorem~\ref{thm:positive_randomized}.
    \end{remark}

\subsection{From Clairvoyant Addition-Robustness to Consistency}
\label{app:rand-checkpoints}

We prove Lemma~\ref{thm:meta} from Section~\ref{sec:rand-robustness} by
adapting the checkpoint construction of \citet{DuttingFLNSZ25}.
Algorithm~\ref{alg:randomized-clairvoyant} retains recent arrivals while
gradually replacing a sampled core. In the algorithm,
\textsc{AdditionRobust} samples from the $\alpha$-addition-robust distribution
supplied by the lemma's hypothesis. The proof verifies feasibility,
bounds both insertions and deletions at every update, and establishes the
expected approximation guarantee at each fixed time.

\begin{algorithm}[H]
\caption{Randomized Clairvoyant Algorithm}
\label{alg:randomized-clairvoyant}
\begin{algorithmic}[1]
\Require Stream $(x_1,\ldots,x_n)$, submodular function $f$,
         budget $k$, parameter $\gamma\in(0,1/8)$
\If{$k<8\gamma^{-2}$}
    \State Maintain an exact optimum after every arrival
    \State \Return
\EndIf
\State $\ell\gets\lfloor\gamma k\rfloor$, $\kappa\gets k-2\ell$,
       $h\gets\lceil 1/\gamma\rceil$
\State Partition the stream into blocks $B_1,\ldots,B_q$ of length $\ell$,
       padding the last block with zero-marginal dummy arrivals if needed
\State $\tau_j\gets j\ell$ for $j=0,\ldots,q$; $A_0\gets\varnothing$
\State During $B_1$, retain all arriving elements
\For{$j=1,\ldots,q-1$}
    \State At boundary $\tau_j$, sample
        $A_j\gets\textsc{AdditionRobust}(X_{\tau_j},(x_{\tau_j+1},\ldots,x_n),f,\kappa)$
    \State $C\gets A_{j-1}\cup B_j$
    \State Pair elements of $C\setminus A_j$ with elements of $A_j\setminus C$
    \State Schedule these exchanges, followed by the remaining deletions or insertions
    \State Partition $B_{j+1}$ into $h$ contiguous subblocks of nearly equal length
    \State Choose a subblock $H$ uniformly at random, independently of checkpoint samples
    \For{each arrival $x_t$ in $B_{j+1}$}
        \If{$x_t\in H$}
            \State Execute the next scheduled operations on $C$, evenly distributed over $H$
        \EndIf
        \State $\mathrm{ALG}_t\gets C\cup(B_{j+1}\cap X_t)$
    \EndFor
\EndFor
\State Discard outputs after the actual end of the stream
\end{algorithmic}
\end{algorithm}
\metatheorem*
\begin{proof}
    For each future prefix $R$, an optimum $T$ of size at most
    $\kappa$ in $V_{\mathrm{now}}\cup R$ satisfies
    $T\subseteq(T\cap V_{\mathrm{now}})\cup R$.
    Hence
    \[
        \max_{\substack{O\subseteq V_{\mathrm{now}}\\|O|\leq\kappa}}
            f(O\cup R)
        \geq
        \max_{\substack{T\subseteq V_{\mathrm{now}}\cup R\\|T|\leq\kappa}}
            f(T).
    \]
    Thus the addition-robust guarantee also bounds the dynamic optimum
    at each prefix, as needed below.

    If $k<8\gamma^{-2}$, recompute an exact optimum after every
    arrival. Its symmetric difference with the previous solution
    has size at most $2k=O(\gamma^{-2})$. Assume henceforth
    $k\geq8\gamma^{-2}$ and set
    \[
        \ell=\lfloor\gamma k\rfloor,\qquad
        \kappa=k-2\ell,\qquad h=\lceil1/\gamma\rceil.
    \]
    Split the stream into blocks $B_1,B_2,\ldots$ of $\ell$ arrivals,
    with boundaries $\tau_j=j\ell$.
    For scheduling only, an incomplete final block may be completed
    by zero-marginal dummy arrivals; outputs after the actual end
    of the stream are discarded.
    At each boundary $\tau_j$, sample a checkpoint solution
    $A_j\subseteq X_{\tau_j}$ of size at most $\kappa$ using the
    clairvoyant routine with the remaining stream. Set $A_0=\varnothing$.
    The first block is handled by retaining all its arrivals.

    \paragraph{Feasibility and recourse.}
    During block $B_{j+1}$, transition from the old core
    $A_{j-1}\cup B_j$ to the new core $A_j$.
    Including $B_j$ in the old core ensures that its
    obsolete elements are deleted as part of this transition,
    rather than all at once at a block boundary.
    Both cores have size at most $\kappa+\ell=k-\ell$.
    Throughout the transition, also retain the current prefix of $B_{j+1}$.
    To change cores, pair elements to be removed with elements
    to be inserted, make the resulting exchanges, and then perform
    any remaining deletions or insertions. There are at most $k$
    such operations; each changes at most two elements, and every
    intermediate core has size at most that of the larger endpoint core.
    The maintained solution therefore always has size at most $k$.

    Partition $B_{j+1}$ into $h$ contiguous subblocks whose lengths
    differ by at most one. Independently of all checkpoint samples,
    choose one subblock uniformly at random and spread the transition
    operations over its updates. The minimum subblock length $d$
    satisfies
    \[
        d=\lfloor\ell/h\rfloor\geq\frac{\gamma^2 k}{8}.
    \]
    Here we use $\ell\geq\gamma k/2$, $h\leq2/\gamma$, and
    $\ell/h\geq2$. Each update needs at most
    $2\lceil k/d\rceil+1=O(\gamma^{-2})$ changes, including the
    new arrival. After the transition the solution is $A_j$ plus
    the current block prefix; at the next boundary this is exactly
    the next old core. Thus there is no additional boundary cost.

    \paragraph{Value at a fixed time.}
    Fix an arrival time $t$ with $x_t\in B_{j+1}$. Its subblock is selected for transition with
    probability $1/h\leq\gamma$. Before the selected subblock,
    the maintained set is
    $A_{j-1}\cup(X_t\setminus X_{\tau_{j-1}})$;
    after it, the set is $A_j\cup(X_t\setminus X_{\tau_j})$.
    Each added set is a prefix of the future sequence supplied to
    the corresponding checkpoint routine. The events determining
    which case occurs are independent of the checkpoint samples.
    Writing $\operatorname{OPT}^{(\kappa)}_t$ for the best solution
    of size at most $\kappa$ in $X_t$, we obtain
    \[
        \mathbb{E}[f(\operatorname{ALG}_t)]
        \geq(1-1/h)\alpha f(\operatorname{OPT}^{(\kappa)}_t).
    \]
    This also covers the first old core $A_0=\varnothing$, since
    its union with the arrivals contains every available element.
    To compare the reduced-budget optimum with $\OPT_t$, suppose first
    that $s=|\OPT_t|\geq\kappa$. By submodularity and nonnegativity, a
    uniformly sampled $\kappa$-subset of $\OPT_t$ has expected value at least
    $(\kappa/s)f(\OPT_t)$, where $\kappa/s\geq\kappa/k$.
    If $\OPT_t$ has fewer than $\kappa$ elements,
    retain it in full. Hence
    \[
        f(\operatorname{OPT}^{(\kappa)}_t)
        \geq\frac{\kappa}{k}f(\operatorname{OPT}_t)
        \geq(1-2\gamma)f(\operatorname{OPT}_t).
    \]
    Combining the last two displays gives an approximation ratio of at least
    $(1-\gamma)(1-2\gamma)\alpha\geq\alpha-3\gamma$.
\end{proof}

\subsection{From Interval Selection to Clairvoyant Addition-Robustness}
\label{app:rand-addition-proofs}

Having established the checkpoint reduction, we turn to the
addition-robust distribution it requires. We use the prefix optimizers
$O_i$, prefixes $R_j$, and LP value $\rho^\star$
from Section~\ref{sec:rand-robustness}. Let $\Delta$ be the simplex of
distributions on $\{0,\ldots,m\}$.

If $\kappa=0$, returning $\varnothing$ achieves the benchmark
$f(R_j)$ exactly for every prefix. If $\kappa\geq1$ and $f(O_0)=0$,
every present element has zero singleton value. Monotonicity and
submodularity then imply $f(O\cup R_j)=f(R_j)$ for every
$O\subseteq V_{\mathrm{now}}$ and every $j$, so returning $O_0$
also achieves the benchmark exactly.

For the remainder of this subsection, assume
$\kappa\geq1$ and $f(O_0)>0$. Define the normalized payoffs
\[
    w_{i,j}=\frac{f(O_i\cup R_j)}{f(O_j\cup R_j)},
\]
which are well defined under these assumptions.
We first justify the minimax formulation used in the main text.

            \begin{lemma}
            \label{lem:minimax}
                Under the standing assumptions $\kappa\geq1$ and $f(O_0)>0$, finite minimax duality gives
                \[
                    \rho^\star =  \max_{\alpha \in \Delta} \min_{j=0,\ldots,m} \sum_{i=0}^m \alpha_i w_{i,j} = \min_{\beta \in \Delta} \max_{i=0, \dots, m} \sum_{j=0}^m \beta_j w_{i,j}.
                \]
            \end{lemma}
            \begin{proof}
                A linear function on a simplex attains its maximum and minimum
                at vertices. Applying this observation before and after finite
                minimax duality gives
                \begin{align*}
                    \max_{\alpha \in \Delta} \min_{j=0,\ldots,m} \sum_i \alpha_i w_{i,j}
                    &= \max_{\alpha \in \Delta} \min_{\beta \in \Delta}
                       \sum_{i=0}^m\sum_{j=0}^m\alpha_i \beta_j w_{i,j}\\
                    &= \min_{\beta \in \Delta}\max_{\alpha \in \Delta}
                       \sum_{i=0}^m\sum_{j=0}^m \alpha_i \beta_j w_{i,j}\\
                    &= \min_{\beta \in \Delta}\max_i
                       \sum_{j=0}^m \beta_j w_{i,j}.
                \end{align*}
            \end{proof}

We now represent each prefix optimizer by an interval. The interval
payoffs lower-bound the corresponding submodular values, and reweighting
the adversary's distribution lets us apply an adaptivity-gap bound to the
normalized payoffs in Lemma~\ref{lem:minimax}.

\lossless*
\begin{proof}
It suffices to consider the nontrivial case described above.
Associate with each $O_i$ the interval
\[
    [\ell_i,r_i]=[f(R_i),f(O_i\cup R_i)].
\]
We first claim that, for every $i,j$,
\begin{equation}
\label{eq:submodular-dominates-interval}
    f(O_i\cup R_j)\geq v(i,j).
\end{equation}
If $j\leq i$, diminishing returns and $R_j\subseteq R_i$ give
\[
    f(O_i\cup R_j)-f(R_j)
    \geq f(O_i\cup R_i)-f(R_i)=r_i-\ell_i.
\]
Since $\ell_j\leq\ell_i$, the right-hand side becomes $v(i,j)$ after
adding $f(R_j)=\ell_j$. If $j\geq i$, monotonicity gives
\[
    f(O_i\cup R_j)\geq\max\{f(R_j),f(O_i\cup R_i)\}
    =\max\{\ell_j,r_i\}=v(i,j).
\]

Fix any $\beta\in\Delta$ and define
\[
    Z=\sum_{j=0}^m\frac{\beta_j}{r_j},
    \qquad
    p_j=\frac{\beta_j/r_j}{Z}\qquad(j=0,\ldots,m).
\]
Then $p$ is a probability distribution on the intervals
$[\ell_j,r_j]$, and $\beta_j=Zp_jr_j$. This reweighting absorbs the
denominators in $w_{i,j}$. By~\eqref{eq:submodular-dominates-interval} and
the assumed bound $\Gamma(\ell,r,p)\geq\alpha$,
\begin{align*}
    \max_i\sum_{j=0}^m\beta_j
        \frac{f(O_i\cup R_j)}{r_j}
    &=Z\max_i\mathbb{E}_{j\sim p}[f(O_i\cup R_j)]\\
    &\geq Z\max_i\mathbb{E}_{j\sim p}[v(i,j)]\\
    &\geq Z\alpha
      \mathbb{E}_{j\sim p}\!\left[\max_h v(h,j)\right]\\
    &\geq Z\alpha\mathbb{E}_{j\sim p}[v(j,j)]\\
    &=Z\alpha\mathbb{E}_{j\sim p}[r_j]
      =\alpha.
\end{align*}
This holds for every $\beta\in\Delta$. Lemma~\ref{lem:minimax} therefore
gives $\rho^\star\geq\alpha$, proving the lossless reduction.
\end{proof}

\subsection{From Structured to General Interval Selection}
\label{sec:rand-interval-proof}

Lemma~\ref{lem:rand-interval-reduction} reduces the algorithmic guarantee to
a uniform interval-selection bound. We next prove
Lemma~\ref{lem:rand-chain-reduction}, which allows us to restrict attention
to the structured family in Section~\ref{sec:rand-hard-structure}.
The proof replaces an arbitrary instance by shadow intervals, normalizes
their expected surviving lengths, and expresses the required inequality
as an average of inequalities for consecutive intervals under equalizing
distributions.

We first derive the equalizing distribution for nonincreasing consecutive
intervals. For every fixed interval $i$,
\[
    \mathbb{E}_{j\sim p}[v(i,j)]
    =\mathbb{E}_{j\sim p}[\ell_j]
     +(r_i-\ell_i)\sum_{j=0}^i p_j.
\]
Since $\sum_{j=0}^N p_j=1$, these values are equal for all $i$ if and only if
\begin{equation}
\label{eq:rand-equalizing-distribution}
    \sum_{j=0}^i p_j=\frac{r_N-\ell_N}{r_i-\ell_i}
    \qquad(i=0,\ldots,N).
\end{equation}
The positive, nonincreasing interval lengths make these cumulative
probabilities nondecreasing and ensure that they end at $1$, so they
define a unique probability distribution. Every fixed interval then has
expected value $\mathbb{E}_{j\sim p}[\ell_j]+(r_N-\ell_N)$.

\chainreduction*
\begin{proof}
The case $\alpha=0$ is immediate. Assume $\alpha>0$ and put
$q_\alpha=(1-\alpha)/\alpha$.
\paragraph{Shadow intervals.}
Write $I_i=[\ell_i,r_i]$. For each realization $j$ with $p_j>0$, set
\[
    d_j=\max_i |I_i\cap[\ell_j,\infty)|,
    \qquad
    \widehat I_j=[\ell_j,\ell_j+d_j],
\]
and choose $i(j)$ attaining the maximum. We call $\widehat I_j$ the
\emph{shadow interval} for realization $j$. If $d_j>0$, the interval
$I_{i(j)}\cap[\ell_j,\infty)$ has length $d_j$ and begins at or to the
right of $\ell_j$. It is therefore a rightward translate of
$\widehat I_j$, so
\begin{equation}
\label{eq:rand-shadow-domination}
    |I_{i(j)}\cap[x,\infty)|
    \geq |\widehat I_j\cap[x,\infty)|
    \qquad\text{for every }x\geq0.
\end{equation}
For $d_j=0$, the same inequality is immediate.
Consequently the best fixed original interval is at least as good as
the best fixed shadow interval. Equation~\eqref{eq:rand-shadow-domination}
also shows that the adaptive value of the shadow instance cannot exceed
that of the original instance. On the other hand, shadow interval
$\widehat I_j$ attains the original hindsight value at realization
$j$, so the two adaptive values are equal, namely
\[
    \sum_j p_j(\ell_j+d_j),
\]
the expected diagonal value of the shadow intervals. Discard
zero-probability realizations and relabel the shadow intervals. It now
suffices to prove the following diagonal bound for every interval family
with $p_i>0$:
\begin{equation}
\label{eq:rand-diagonal-form}
    \alpha\mathbb{E}_{j\sim p}[r_j]
    \leq
    \mathbb{E}_{j\sim p}[\ell_j]
    +\max_i\mathbb{E}_{j\sim p}
       \bigl[|I_i\cap[\ell_j,\infty)|\bigr].
\end{equation}

\paragraph{Normalizing the fixed value.}
Let
\[
    B=\mathbb{E}_{j\sim p}[\ell_j],\qquad
    C=\mathbb{E}_{j\sim p}[r_j-\ell_j],\qquad
    M=\max_i\mathbb{E}_{j\sim p}
       \bigl[|I_i\cap[\ell_j,\infty)|\bigr].
\]
Thus~\eqref{eq:rand-diagonal-form} is $B+M\geq\alpha(B+C)$.
Subtracting $\min_i\ell_i$ from every endpoint leaves $M$ and $C$
unchanged. Restoring this translation adds
$(1-\alpha)\min_i\ell_i\geq0$ to the slack, so we assume
$\min_i\ell_i=0$.

Define the distribution function of the random left endpoint by
\[
    W(x)=\Pr_{j\sim p}[\ell_j\leq x].
\]
Fubini's theorem gives
\begin{equation}
\label{eq:rand-interval-area}
    M=\max_i\int_{\ell_i}^{r_i}W(x)\,dx.
\end{equation}
If $M=0$, then
$p_i(r_i-\ell_i)\leq\int_{\ell_i}^{r_i}W(x)\,dx=0$ for every $i$,
so $C=0$ and the claim follows. Assume $M>0$. By homogeneity, divide
all endpoints by $M$ and henceforth suppose $M=1$.

Extend each interval to the right until
\begin{equation}
\label{eq:rand-saturation}
    \int_{\ell_i}^{r_i}W(x)\,dx=1
    \qquad\text{for every }i.
\end{equation}
Such extensions exist because $W(x)\geq W(0)>0$. They increase $C$
while keeping $B$ and $M=1$ fixed, so it is enough to prove the
stronger inequality for the extended instance.

\paragraph{A change of coordinates.}
The saturation step gives every interval the same area under $W$. We use
this area as a new coordinate, in which every interval has length $1$.
Set
\[
    H(x)=\int_0^xW(t)\,dt,\qquad X=H^{-1},\qquad
    g(y)=\frac{1}{W(X(y))}.
\]
Then $g$ is nonincreasing, $g\geq1$, and $g$ is eventually equal to
$1$; moreover, $X'(y)=g(y)$ almost everywhere. The tail-integral
formula and the substitution $y=H(x)$ yield
\begin{equation}
\label{eq:rand-B}
    B=\int_0^\infty(1-W(x))\,dx
     =\int_0^\infty(g(y)-1)\,dy.
\end{equation}
If $J\sim p$ and $S=H(\ell_J)$, then~\eqref{eq:rand-saturation}
implies
\[
    r_J-\ell_J
    =X(S+1)-X(S)=\int_S^{S+1}g(y)\,dy.
\]
For $y\geq0$,
\[
    \Pr[S\leq y]=\Pr[\ell_J\leq X(y)]
    =W(X(y))=\frac{1}{g(y)}.
\]
To average these lengths, integrate $g(y)\Pr[S\leq y\leq S+1]$ over
$y\geq0$. The contribution from $[0,1)$ is $1$; shifting the remaining
integral by $1$ gives, with values at jump points irrelevant,
\begin{equation}
\label{eq:rand-C}
    C=1+\int_0^\infty
       \left(1-\frac{g(y+1)}{g(y)}\right)\,dy.
\end{equation}

\paragraph{Averaging consecutive instances.}
The two integral formulas now connect the arbitrary instance to the
structured family. Choose an integer $N$ such that $g(y)=1$ for $y\geq N$.
For every
$s\in[0,1)$, form consecutive intervals of lengths
$g(s),g(s+1),\ldots,g(s+N)=1$, starting at zero, and equip them
with their equalizing distribution. Using the cumulative probabilities
in~\eqref{eq:rand-equalizing-distribution}, their best fixed expected
value is $1+\sum_{j=0}^{N-1}(g(s+j)-1)$, and their additional
expected value from adaptation is
$\sum_{j=0}^{N-1}(1-g(s+j+1)/g(s+j))$. Rearranging the assumed
bound $\Gamma(\ell,r,p)\geq\alpha$ for these instances gives
\[
    \sum_{j=0}^{N-1}\left(1-\frac{g(s+j+1)}{g(s+j)}\right)
    \leq q_\alpha\left(1+\sum_{j=0}^{N-1}(g(s+j)-1)\right).
\]
Integrating over $s\in[0,1)$ and using
\eqref{eq:rand-B} and~\eqref{eq:rand-C} gives
\[
    C-1\leq q_\alpha(1+B).
\]
Since $\alpha=1/(1+q_\alpha)$,
\[
    \alpha(B+C)\leq B+1=B+M.
\]
Undoing the extensions, scaling, and translation proves the diagonal
bound~\eqref{eq:rand-diagonal-form}. The shadow-interval comparison then
gives $\Gamma(\ell,r,p)\geq\alpha$ for the original instance, as required.
\end{proof}

\subsection{The Adaptivity Gap for Structured Interval Selection}
\label{app:rand-structured}

By Lemma~\ref{lem:rand-chain-reduction}, a sharp bound for consecutive
intervals with the equalizing distribution suffices for all interval
instances. We prove Lemma~\ref{lem:rand-chain} by expressing this bound as
an inequality in the interval lengths. Optimizing a chain of three
intervals identifies the parameters $u_1,u_2$ used in the coverage
construction; we then show that longer chains cannot give a smaller
adaptivity gap.

\sharpadaptive*
\begin{proof}
    Put $q=(1-\xi^\star)/\xi^\star$. Equivalently, $q$ is the unique
    root in $(0,1)$ of
    \begin{equation}
    \label{eq:rand-Q}
        Q(q)=q^4+q^3-22q^2-87q+24=0.
    \end{equation}
    This characterization follows from
    $Q(t)=(1+t)^4\Psi(1/(1+t))$,
    $Q(0)>0>Q(1)$, and $Q'(t)<0$ on $(0,1)$.
    Evaluation at two rational points also gives
    \begin{equation}
    \label{eq:rand-q-range}
        \frac{259}{1000}<q<\frac{13}{50}.
    \end{equation}

    By scaling all endpoints, assume $r_N-\ell_N=1$.
    Set $d_j=r_j-\ell_j$ for $j=0,\ldots,N$.
    Then $d_0\geq\cdots\geq d_N=1$, and~\eqref{eq:rand-equalizing-distribution} gives
    $\sum_{i=0}^j p_i=1/d_j$.
    For each fixed interval $i$,
    \[
        \mathbb{E}_{j\sim p}
        \bigl[|[\ell_i,r_i]\cap[\ell_j,\infty)|\bigr]
        =d_i\sum_{j=0}^i p_j=1.
    \]
    For realization $j$, interval $j$ is a best interval in hindsight:
    earlier intervals have no surviving length, and later intervals
    have length at most $d_j$. Hence the best fixed expected value is
    $B+1$ and the adaptive expected value is $B+C$, where summation
    by parts gives
    \[
        B=\sum_{j=0}^N p_j\ell_j
          =\sum_{j=0}^{N-1}(d_j-1),
        \qquad
        C=\sum_{j=0}^N p_jd_j
          =1+\sum_{j=0}^{N-1}\left(1-\frac{d_{j+1}}{d_j}\right).
    \]
    Conversely, every finite chain $d_0\geq\cdots\geq d_N=1$
    defines such an interval instance. Thus the claimed bound is
    equivalent to the numerical inequality
    \begin{equation}
    \label{eq:rand-chain}
        \sum_{j=0}^{N-1}\left(1-\frac{d_{j+1}}{d_j}\right)
        \leq q\left(1+\sum_{j=0}^{N-1}(d_j-1)\right).
    \end{equation}
    Define
    \[
        \varphi(x,y)=1-y/x-q(x-1)\qquad(x\geq y\geq1).
    \]
    The required inequality is equivalent to
    $\sum_{j=0}^{N-1}\varphi(d_j,d_{j+1})\leq q$.

    \paragraph{Two-edge chains: three intervals.}
    We first optimize over chains $x\to y\to1$, which correspond to
    three consecutive intervals. The smallest coefficient for which
    \eqref{eq:rand-chain} holds for all such chains is the maximum of
    \[
        R(x,y)=\frac{2-y/x-1/y}{x+y-1},\qquad x\geq y\geq1.
    \]
    On the boundary $y=1$,
    $R(x,1)=(x-1)/x^2\leq1/4$; on $y=x$,
    \[
        R(x,x)=\frac{1-1/x}{2x-1}
        \leq\frac{x-1}{x^2}\leq\frac14.
    \]
    Also $R(x,y)\leq2/x$, whereas $R(2,6/5)=17/66>1/4$.
    Hence its maximum $q_0$ is attained at an interior point
    $x>y>1$. The first-order conditions give
    \[
        q_0=\frac{y}{x^2}=\frac{1}{y^2}-\frac{1}{x}.
    \]
    These conditions and $R(x,y)=q_0$ imply
    \[
        \Phi(y/x)=0,\qquad
        \frac1x=\left(\frac yx\right)^2\left(1+\frac yx\right).
    \]
    The polynomial $\Phi$ has exactly one root in $(0,1)$ because
    $\Phi(0)=2$, $\Phi(1)=-1$, and
    $\Phi'(t)=(t^2-1)(4t+3)<0$ for $t\in(0,1)$.
    Since $0<y/x<1$, uniqueness gives $y/x=u_1$, and the definition
    of $u_2$ gives $1/x=u_2$. Consequently,
    $q_0=u_1u_2=2u_1^2+3u_1-2$. The identity
    \[
        Q(2t^2+3t-2)
        =\Phi(t)(16t^4+80t^3+112t^2+60t+59)
    \]
    evaluated at $t=u_1$ shows that $Q(q_0)=0$.
    Since $q_0=u_1u_2\in(0,1)$, uniqueness of the root of $Q$
    gives $q_0=q$. Thus
    \begin{equation}
    \label{eq:rand-two-edge}
        \varphi(x,y)+\varphi(y,1)\leq q\qquad(x\geq y\geq1),
    \end{equation}
    with equality at $x=1/u_2$, $y=u_1/u_2$. The interior maximizer also
    implies $0<u_2<u_1<1$. A one-edge chain is covered by $y=1$.

    \paragraph{Controlling chains of arbitrary length.}
    We next reduce longer chains to the case just analyzed.
    If an edge $d_j\to d_{j+1}$ has nonpositive $\varphi$-value,
    delete its starting vertex. Deleting the first vertex removes
    a nonpositive term. Otherwise the preceding edge changes from
    $d_{j-1}\to d_j$ to $d_{j-1}\to d_{j+1}$, increasing its value
    by $(d_j-d_{j+1})/d_{j-1}\geq0$.
    These deletions cannot decrease the total $\varphi$-value.
    We may therefore restrict attention to chains whose edges all
    have positive value. A chain with no edges has value zero, and
    chains with one or two edges satisfy~\eqref{eq:rand-two-edge}.
    Assume henceforth that the remaining chain has at least three edges.

    A positive edge $x\to y$ satisfies
    \[
        y<h(x):=(1+q)x-qx^2
        \leq T:=\frac{(1+q)^2}{4q}.
    \]
    The bounds in~\eqref{eq:rand-q-range} imply $T<31/20$ and
    that $h$ is increasing on $[1,31/20]$. Hence
    \[
        d_1<T,\qquad
        d_2<h(d_1)<h(31/20)
        =\frac{31}{20}-\frac{341q}{400}<\frac43.
    \]
    Set $L=1/(1-q)>4/3$. Whenever $1\leq z\leq y\leq x\leq L$,
    \begin{align}
        \varphi(x,y)+\varphi(y,z)-\varphi(x,z)
        &=(y-z)\left(\frac1y-\frac1x\right)-q(y-1)\notag\\
        &\leq(y-1)\left(1-\frac1x-q\right)\leq0.
        \label{eq:rand-collapse}
    \end{align}
    Thus the tail starting at $d_2$ can be collapsed to a single
    edge ending at $1$. It remains to bound three-edge chains
    $x\to y\to z\to1$. If $y\leq L$, another application
    of~\eqref{eq:rand-collapse} reduces to two edges.

    Suppose $y\geq L$. The arithmetic--geometric mean inequality gives
    \begin{align*}
        \varphi(x,y)+\varphi(y,z)+\varphi(z,1)
        &=3+3q-qy-\left(qx+\frac yx\right)
          -\left(\left(q+\frac1y\right)z+\frac1z\right)\\
        &\leq F(y):=3+3q-qy-2\sqrt{qy}-2\sqrt{q+1/y}.
    \end{align*}
    For $y\geq L>4/3$,
    \[
        F'(y)=-q-\sqrt{q/y}+\frac{1}{y^2\sqrt{q+1/y}}
        \leq-q-\sqrt{q/y}+y^{-3/2}<0.
    \]
    Indeed, $q>1/4$ implies
    \[
        qy^{3/2}+\sqrt q\,y
        >\frac14\left(\frac43\right)^{3/2}
         +\frac12\cdot\frac43
        =\frac{2}{3\sqrt3}+\frac23>1.
    \]
    Since $q+1/L=1$ and $q<1/2$, the point
    $(x,y,z)=(\sqrt{L/q},L,1)$ is admissible and attains equality
    in both arithmetic--geometric mean inequalities. Therefore
    \[
        F(y)\leq F(L)
        =\varphi(\sqrt{L/q},L)+\varphi(L,1)\leq q
    \]
    by~\eqref{eq:rand-two-edge}. This proves the inequality for all
    finite chains. Equality holds for
    \begin{equation}
    \label{eq:rand-extremal-chain}
        d_0=1/u_2,\qquad d_1=u_1/u_2,\qquad d_2=1.
    \end{equation}
    The corresponding consecutive intervals, after scaling all
    endpoints by $u_2$, are precisely those in
    Lemma~\ref{lem:rand-chain}, with the stated probabilities. The equalizing
    probabilities are also the weights $\mu_0,\mu_1,\mu_2$ used in the impossibility proof
    in Appendix~\ref{app:rand-lower-bound}.
    This proves both the bound and its equality case.
\end{proof}

\subsection{Combining the Reductions: The Randomized Approximation Guarantee}
\label{app:rand-combination}

All ingredients of the roadmap in Figure~\ref{fig:rand-reductions} are now
in place. The interval-selection bound supplies the addition-robust
routine assumed by the checkpoint reduction. We finish by choosing its
accuracy parameter to obtain the guarantee stated in
Theorem~\ref{thm:positive_randomized}.

\positiverand*
\begin{proof}
By Lemma~\ref{lem:rand-chain}, the adaptivity gap is at least
$\xi^\star$ for every instance of nonincreasing consecutive
intervals with the equalizing distribution.
Lemma~\ref{lem:rand-chain-reduction} extends this bound to every
interval-selection instance.
Applying Lemma~\ref{lem:rand-interval-reduction} with
$\alpha=\xi^\star$ therefore yields, for every clairvoyant
addition-robustness instance, a $\xi^\star$-addition-robust
distribution.
Use an optimal solution to LP~\eqref{lp:randomized} to sample from this
distribution, and apply Lemma~\ref{thm:meta} with
$\gamma=\min\{\varepsilon/3,1/16\}$.
The resulting randomized clairvoyant algorithm achieves an
approximation ratio of at least
$\xi^\star-3\gamma\geq\xi^\star-\varepsilon$
and is $O(\gamma^{-2})=O(\varepsilon^{-2})$-consistent.
\end{proof}
\section{Proofs for the Tight Deterministic Bound}
\label{app:sharp-deterministic}

This appendix supplies the details deferred from
Section~\ref{sec:sharp-deterministic}. Appendix~\ref{app:det-impossibility}
makes the rows-to-columns argument of
Section~\ref{sec:det-impossibility} precise, including the losses from
bounded recourse and partially arrived stages.
Appendix~\ref{d2:sec:deterministic-two} follows the certificate framework
in Section~\ref{sec:det-framework}:
Appendix~\ref{d2:sec:local} proves the mixture comparisons,
Appendix~\ref{d2:sec:half} gives the elementary certificate construction, and
Appendices~\ref{d2:app:rounding} and~\ref{d2:app:schedule} prove the
reduction to consistent algorithms. The three-quarter certificate proof is
supplied separately, as indicated after Theorem~\ref{thm:det_threequarters}.

\subsection{Impossibility for Deterministic Algorithms}
\label{app:det-impossibility}

In this section, we present an impossibility result holding for deterministic algorithms. In particular, we construct a coverage function and a stream of elements such that any $\eps k$-consistent algorithm cannot achieve more than a $\nicefrac 34+O(\eps)$ approximation of the optimum. More formally, we have the following result, which is proved in the rest of the section.

\impossibilitydeterministic*

\subsubsection{Instance Construction}

Consider the universe set $U$ composed by two disjoint sets: the ``core'' $\Ucore$ and the ``auxiliary'' part $\Uaux$. More precisely, the set $U$ is composed by
\begin{itemize}
    \item $\Ucore = \{1, 2, \dots, k^2\}$ consists of exactly $k^2$ items, which form a $k \times k$ grid.
    \item $\Uaux = \{u_{i,j} \mid i \in \{1, \dots, k\},\; j \in \{1, \dots, m\} \}$ consists of $km$ distinct, newly introduced unique items.
\end{itemize}
The total size of the universe is thus $|U| = k^2 + km$. We introduce the $\cU$ family of subsets of $U$ which are used in the stream. There are two types of subsets: $A$ and $B$.

\paragraph{Construction of $A$.} The elements in $A$, indexed by $\ell \in \{1, \dots, k\}$ covers $\Ucore$ ``by rows'': for each $\ell$, the element $a_\ell$ covers the $\ell$-th block of $k$ items: $a_\ell= \{(\ell-1)k + 1, (\ell-1)k + 2, \dots, \ell k\}$.

By this construction, the sets $a_\ell$ are pairwise disjoint and perfectly cover $\Ucore$. Selecting all $k$ elements from $A$ yields a coverage of $f(A) = k^2$. Note that $a_\ell \cap \Uaux = \emptyset$ for all  $\ell \in \{1, \dots, k\}$.

\paragraph{Construction of $B$.} Family $B$ is further partitioned into $m$ distinct sequences $B_1,B_2, \dots, B_m$, and the sets in a given column are \textit{nested} across these sequences. More precisely, $B_j$ is composed by elements $b_{i,j}$, one for each column index $i \in \{1, \dots, k\}$. In particular, the generic element $b_{i,j}$ covers all the elements in the $i^{th}$ column of $\Ucore$ plus the elements in $\Uaux$ of the type $\{u_{i,1},\dots, u_{i,j}\}$. By design, we then have that the total cardinality of an extended set grows linearly with $j$: $|b_{i,j}| = k + j$.

\begin{figure}[t!]
    \centering
    \resizebox{0.65\textwidth}{!}{%
        \begin{tikzpicture}[
                coreitem/.style={circle, draw=black!80, thick, fill=white, minimum size=12mm, inner sep=0pt, font=\sffamily\small\bfseries, align=center},
                auxitem/.style={circle, draw=black!80, thick, densely dashed, fill=black!5, minimum size=12mm, inner sep=0pt, font=\sffamily\normalsize},
                dotsnode/.style={minimum size=12mm, inner sep=0pt, font=\Huge, align=center},
                %
                setA/.style={draw=blue!80, very thick, rounded corners=15pt, fill=blue!20, fill opacity=0.35},
                nest1/.style={draw=red!80, thick, dotted, rounded corners=15pt},
                nest2/.style={draw=red!90, thick, dashed, rounded corners=22pt},
                nest3/.style={draw=red, very thick, rounded corners=30pt, fill=red!20, fill opacity=0.35}
            ]
            %
            \node[coreitem] (c11) at (0, 0) {$1$};
            \node[coreitem] (c12) at (5.5, 0) {$2$};
            \node[dotsnode] (c1d) at (11, 0) {$\cdots$};
            \node[coreitem] (c1k) at (16.5, 0) {$k$};
            %
            \node[coreitem] (c21) at (0, -3) {$k\!+\!1$};
            \node[coreitem] (c22) at (5.5, -3) {$k\!+\!2$};
            \node[dotsnode] (c2d) at (11, -3) {$\cdots$};
            \node[coreitem] (c2k) at (16.5, -3) {$2k$};
            %
            \node[dotsnode] (v1)  at (0, -6) {$\vdots$};
            \node[dotsnode] (v2)  at (5.5, -6) {$\vdots$};
            \node[dotsnode] (vd)  at (11, -6) {$\ddots$};
            \node[dotsnode] (vk)  at (16.5, -6) {$\vdots$};
            %
            \node[coreitem] (ck1) at (0, -9) {$(k\!-\!1)k$\\$+1$};
            \node[coreitem] (ck2) at (5.5, -9) {$(k\!-\!1)k$\\$+2$};
            \node[dotsnode] (ckd) at (11, -9) {$\cdots$};
            \node[coreitem] (ckk) at (16.5, -9) {$k^2$};
            %
            \draw[black!50, very thick, loosely dashed] (-3, -10.75) -- (19.5, -10.75);
            \node[font=\sffamily\small, text=black, anchor=south west] at (-3, -10.75) {$U_{\text{core}}$ (Arbitrary $k \times k$ Core Grid)};
            \node[font=\sffamily\small, text=black, anchor=north west] at (-3, -10.75) {$U_{\text{aux}}$ (Globally Unique Extensions up to depth $m$)};
            %
            \node[auxitem] (u12) at (0, -12.5) {$u_{1,1}$};
            \node[auxitem] (u22) at (5.5, -12.5) {$u_{2,1}$};
            \node[dotsnode] (ud2) at (11, -12.5) {$\cdots$};
            \node[auxitem] (uk2) at (16.5, -12.5) {$u_{k,1}$};
            %
            \node[dotsnode] (vu1) at (0, -14.5) {$\vdots$};
            \node[dotsnode] (vu2) at (5.5, -14.5) {$\vdots$};
            \node[dotsnode] (vud) at (11, -14.5) {$\ddots$};
            \node[dotsnode] (vuk) at (16.5, -14.5) {$\vdots$};
            %
            \node[auxitem] (u1m) at (0, -16.5) {$u_{1,m}$};
            \node[auxitem] (u2m) at (5.5, -16.5) {$u_{2,m}$};
            \node[dotsnode] (udm) at (11, -16.5) {$\cdots$};
            \node[auxitem] (ukm) at (16.5, -16.5) {$u_{k,m}$};
            %
            \node[text=blue!90!black, font=\bfseries\Huge] at ($(v1)!0.5!(v2)+(-2.8,0)$) {$\vdots$};
            \node[text=red!90!black, font=\bfseries\Huge] at ($(vd)!0.5!(vud)+(0,-1.75)$) {$\cdots$};
            %
            \begin{scope}[on background layer]
                \node[setA, fit=(c11) (c12) (c1d) (c1k), inner sep=8pt] (boxA1) {};
                \node[setA, fit=(c21) (c22) (c2d) (c2k), inner sep=8pt] (boxA2) {};
                \node[setA, fit=(ck1) (ck2) (ckd) (ckk), inner sep=8pt] (boxAk) {};
                %
                \node[nest2, fit=(c11) (c21) (v1) (ck1) (u12), inner sep=24pt] (boxB12) {};
                \node[nest3, fit=(c11) (c21) (v1) (ck1) (u12) (vu1) (u1m), inner sep=32pt] (boxB1m) {};
                %
                \node[nest3, fit=(c12) (c22) (v2) (ck2) (u22) (vu2) (u2m), inner sep=32pt] (boxB2m) {};
                %
                \node[nest3, fit=(c1k) (c2k) (vk) (ckk) (uk2) (vuk) (ukm), inner sep=32pt] (boxBkm) {};
            \end{scope}
            %
            \node[text=blue!90!black, font=\bfseries\Large, anchor=east] at ([xshift=-8pt]boxA1.west) {$a_1$};
            \node[text=blue!90!black, font=\bfseries\Large, anchor=east] at ([xshift=-8pt]boxA2.west) {$a_2$};
            \node[text=blue!90!black, font=\bfseries\Large, anchor=east] at ([xshift=-8pt]boxAk.west) {$a_k$};
            \node[text=red!80!black, font=\bfseries, anchor=west] at ([xshift=6pt]boxB12.east |- u12) {$b_{1,1}$};
            \node[text=red!90!black, font=\bfseries\Large, anchor=north] at ([yshift=-8pt]boxB1m.south) {$b_{1,m}$};
            \node[text=red!90!black, font=\bfseries\Large, anchor=north] at ([yshift=-8pt]boxB2m.south) {$b_{2,m}$};
            \node[text=red!90!black, font=\bfseries\Large, anchor=north] at ([yshift=-8pt]boxBkm.south) {$b_{k,m}$};
            \node[text=red!90!black, font=\bfseries\Huge] at ($(boxB2m.south)!0.5!(boxBkm.south)+(0,-0.3)$) {$\cdots$};
        \end{tikzpicture}%
    }
    \caption{Construction of the hard instance for deterministic consistent algorithms.}
    \label{fig:construction}
\end{figure}
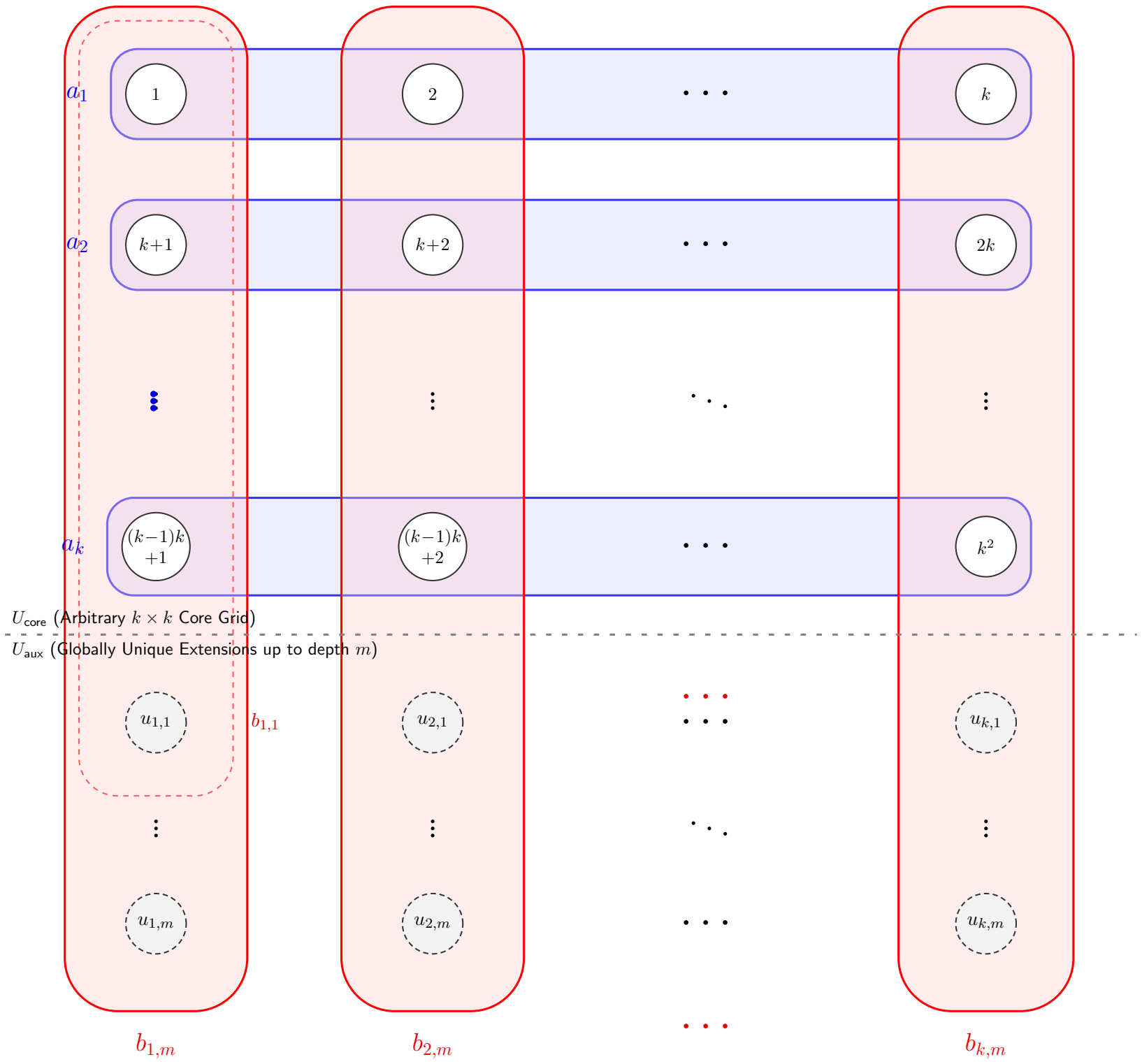

Note, any algorithmic strategy that attempts to mix elements from $A$ and elements from $B$ inevitably incurs heavy redundantly-covered items within the core grid. Meanwhile, the unique extension elements $u_{i,j}$ allow an adversary to finely tune the optimal values of the unpicked $b$-chains. For a visualization, we refer to \Cref{fig:construction}.

\paragraph{Streaming Arrival Sequence.} The elements in $A$ and $B$ are presented to the algorithm into $m+1$ chronological phases. The stream $\Sigma$ begins by presenting the entire collection $A$, revealing the core elements $a_1, a_2, \dots, a_k$. Following $A$, the nested elements of collection $B$ arrive in $m$ distinct stages, ordered by their sequence depth $j$. Specifically, for each stage $j$ from $1$ to $m$, the stream sequentially presents the $k$ elements $\langle b_{1,j}, b_{2,j}, \dots, b_{k,j} \rangle$ associated with that depth. Formally, letting $B_j = \langle b_{1,j}, \dots, b_{k,j} \rangle$, the complete adversarial sequence is defined by the concatenation of these phases:
\[
    \Sigma = A \circ B_1 \circ B_2 \circ \dots \circ B_m
\]
The stream is fixed and may be known to the algorithm in advance; the argument below only relies on how slowly a consistent solution can change.

\subsubsection{Algorithms behavior on stream \texorpdfstring{$\Sigma$}{Sigma}}

Fix any deterministic (and possibly clairvoyant) algorithm for the problem. To maintain a tight competitive ratio any algorithm is forced into a stringent set of mutually exclusive choices regarding the retention of early elements. Denote with $n_t(\ALG)=n_t$ the number of elements in $A$ that the algorithm maintains in its solution at the end of round $t$ (i.e., after the computation following the insertion of the $t^{th}$ element of the stream). Since the algorithm and the stream are deterministic, the $n_t$ form a deterministic sequence. We have the following lemma.

\begin{lemma}
    \label{lem:intermediate}
    Let \ALG be any deterministic algorithm for the Maximum Coverage Problem subject to a uniform cardinality constraint $k$ which maintains a $\nicefrac 34$ approximation to the dynamic optimum, then:
    \begin{enumerate}
        \item $n_k(\ALG) \ge \frac{3}{4}k$.
        \item There exists a $t>k$ such that $n_t(\ALG) \le \frac{1}{3}k$, provided that $m \ge k$.
    \end{enumerate}
\end{lemma}
\begin{proof}
    The proof of the first property is immediate: the dynamic optimum after the $k^{th}$ insertion covers all $\Ucore$, which contains $k^2$ elements, while the elements covered by the algorithm is exactly $n_k \cdot k$ (as each element in $A$ covers exactly a distinct row of $k$ elements in $\Ucore$).

    Consider now a generic time step $t$ of the form $k \times (j+1)$, for $j\ge 1$, i.e., the end of stage $j$. In that moment the optimal solution consists of the last $k$ elements inserted, i.e., in $B_j$, which cover $k^2+ k\times j$ universe elements: these sets are disjoint, and no set inserted so far has more than $k+j$ universe elements. Conversely, the value of the algorithm is
    \begin{align*}
        f(\ALG_t) & \le k \cdot n_t(\ALG) + (k -n_t(\ALG)) \cdot (j + k)-n_t(\ALG)(k-n_t(\ALG)) \\
                  & = n_t^2 + (k-n_t)(j+k)
    \end{align*}

    Indeed, each one of the $n_t$ elements in $A$ maintained by the algorithm covers exactly $k$ elements, and the solution has at most $(k-n_t)$ elements from $B$. For an upper bound, these can be replaced by exactly $k-n_t$ elements of $B_j$ from distinct columns, whether or not all of them have arrived: an element from an earlier stage is contained in the element of $B_j$ in its column, repeated columns are dominated by the latest set in that column, and unused slots can be filled with other columns. Each of these replacements covers $j$ elements in $\Uaux$ and $k$ elements in $\Ucore$, and there are exactly $n_t \cdot (k-n_t)$ elements in $\Ucore$ which are covered by exactly one element from $A$ and one from $B$. This bound holds at every time step of stage $j$, not only at its end.

    Putting everything together, we have:
    \[
        \frac 34 \le \frac{f(\ALG_t)}{f(\OPT_t)} \le \frac{n_t^2 + (k-n_t)(j+k)}{k^2+ k\times j}
    \]

    For simplicity, set $j = k$, and let $\alpha_t$ denote the \emph{fraction} $\nicefrac{n_t}{k}.$ The above inequality becomes:
    \[
        4 \alpha_t^2 - 8 \alpha_t + 2 \ge 0.
    \]
    The fraction $\alpha_t$ is at most one, thus solving the inequality we get that $\alpha_t$ is at most $1 - \nicefrac{\sqrt 2}{2} \le \nicefrac 13$. This concludes the proof.
\end{proof}

\begin{remark}
    Note, the constant $\nicefrac 13$ in the statement of the lemma is not important in itself. We only want to argue, via an intermediate point value argument, that the solution maintained by the algorithm contains approximately half of the elements in $A$.
\end{remark}

We have all the ingredients to prove the main theorem of the section.

\begin{proof}[Proof of \Cref{thm:det_ub}]
    Consider the instance constructed so far with $m=k$ and any deterministic algorithm which is $\eps k$ consistent. We may assume that $\eps<\nicefrac 14$, as otherwise the claim follows from the trivial upper bound of $1$ on the approximation ratio. We may also assume that the algorithm maintains a $\nicefrac 34$ approximation at all times, as otherwise the claim already holds. Consistency gives $|n_t-n_{t-1}| \le |\ALG_t \triangle \ALG_{t-1}| \le \eps k$. By an intermediate point argument and \Cref{lem:intermediate} there exists some time step $t \ge k$ such that $\nicefrac{n_t}k \in [\nicefrac 12,\nicefrac 12(1+2\eps)]$, and $t>k$ since $n_k \ge \nicefrac 34 k > (\nicefrac 12+\eps)k$. Such time step $t$ will correspond to a batch $B_j$, for some $j \ge 1$, where the optimal solution is at least $k \times (k+j-1)$, as all of $B_{j-1}$ (or all of $A$, if $j=1$) has already arrived. Conversely, by the argument in the proof of \Cref{lem:intermediate}, the solution maintained by the algorithm has value at most:
    \begin{align*}
        f(\ALG_t) & \le n_tk + (k-n_t) (k+j) - n_t \cdot (k-n_t)  \\
                  & \leq k^2 + kj - n_tj -n_tk+n_t^2              \\
                  & \leq k^2 + kj -kj/2-k^2/2 + (k+2\eps k)^2/4   \\
                  & \leq \frac{3}{4}(k+2\eps k)^2 + \frac{1}{2}kj \\
    \end{align*}
    Therefore we get that
    \[
        \frac{f(\ALG_t)}{f(\OPT_t)} \le \frac{\frac{3}{4}(k+2\eps k)^2 + \frac{1}{2}kj}{k \times (k+j-1)} \leq \frac{3}{4}(1+2\eps)^2\,\frac{k+j}{k+j-1} \leq 3/4+O(\eps+1/k) \leq 3/4+O(\eps)\,,
    \]
    where the third inequality uses $j \ge 1$, and the last inequality follows from the assumption of the \Cref{thm:det_ub}.
\end{proof}

\subsection{Deterministic Clairvoyant Algorithms}
\label{d2:sec:deterministic-two}
Recall the window instances, mixtures, and certificates from
Section~\ref{sec:det-framework}. We normalize
$f(\varnothing)=0$ throughout the proof; restoring a nonnegative
baseline only improves the guarantees.

\subsubsection{Window bounds and mixture comparisons}
\label{d2:sec:local}

\paragraph{Window bounds.}
Fix a window instance with stages $0,\ldots,m$, core budget $\kappa\ge1$,
and candidates $O_0,\ldots,O_m$, with the initial and terminal choices
prescribed. We may keep only selected prefixes as stages, so consecutive
buffers can differ by several arrivals. Set
\begin{equation}\label{d2:eq:benchmarks}
    \lambda_j=f(R_j),\qquad
    P_j=f(O_j\cup R_j),\qquad
    g_j=P_j-\lambda_j,\qquad V=P_0.
\end{equation}
Recall that $P_j$ is the benchmark~\eqref{eq:det-benchmark} at stage $j$,
that is, the value attainable by the entire buffer and an optimal
old core. Different labels may denote
the same physical set.

\begin{dettwolemma}[Window bounds]\label{d2:lem:window-bounds}
    The values $\lambda_j$ and $P_j$ are nondecreasing in $j$, the
    residual $g_j$ is nonincreasing in $j$, and for all stages $s\le j$,
    \begin{equation}\label{d2:eq:basic-facts}
        0\le g_j\le V\le P_j,\qquad
        0\le P_j-P_s\le\lambda_j-\lambda_s.
    \end{equation}
\end{dettwolemma}
\begin{proof}
    Monotonicity makes $\lambda_j$ nondecreasing.
    Fix $s\le j$. Since $R_s\subseteq R_j$
    and $O_j\cap R_j=\varnothing$, submodularity applied to
    $R_s\cup O_j$ and $R_j$ gives
    \[
        f(R_s\cup O_j)+f(R_j)
        \ge f(R_j\cup O_j)+f(R_s).
    \]
    Crucially, the feasible family of old cores is fixed across stages,
    so $O_j$ is also feasible in the maximization~\eqref{eq:det-benchmark} at stage $s$.
    Consequently,
    \begin{align*}
        P_j-P_s
         & \le f(R_j\cup O_j)-f(R_s\cup O_j) \\
         & \le f(R_j)-f(R_s)
        =\lambda_j-\lambda_s.
    \end{align*}
    Monotonicity also gives $P_j\ge P_s$, and hence
    \[
        0\le P_j-P_s\le\lambda_j-\lambda_s.
    \]
    Rearranging the upper bound shows that
    \[
        g_j=P_j-\lambda_j\le P_s-\lambda_s=g_s.
    \]
    Finally, $g_j\ge0$ because the empty core is feasible,
    $g_0=P_0=V$ because $R_0=\varnothing$ and $f(\varnothing)=0$,
    and $P_j\ge P_0=V$ by monotonicity. Therefore,
    \[
        0\le g_j\le V\le P_j.
    \]
\end{proof}

If $V=0$, every old set has value zero and $P_j=\lambda_j$; the buffer
alone suffices. Below we assume $V>0$.

\paragraph{Type and physical mixtures.}
The physical fractional solutions lie in the cardinality polytope
\[
    \mathcal P_\kappa
    =
    \left\{z\in[0,1]^{V_{\mathrm{now}}}:\sum_{e\in V_{\mathrm{now}}}z_e\le\kappa\right\}.
\]
For a buffer $Q$, define the multilinear extension
\[
    F_Q(z)=\dTwoE\bigl[f(Q\cup S(z))\bigr],
\]
where $S(z)\subseteq V_{\mathrm{now}}$ includes each physical element $e$
independently with probability $z_e$. An integral core $A$ is
represented by its indicator vector $\dTwoOne_A$, for which
$F_Q(\dTwoOne_A)=f(Q\cup A)$.

We restrict attention to convex combinations of the candidates
$O_0,\ldots,O_m$, treating their labels as \emph{types}. Recall that
a mixture is a point $p\in\Delta$.
It determines the physical fractional vector
\begin{equation}\label{d2:eq:physical-vector}
    z(p)=\sum_{i=0}^m p_i\dTwoOne_{O_i},
    \qquad
    z_e(p)=\sum_{i:e\in O_i}p_i.
\end{equation}
This vector is feasible because $0\le z_e(p)\le1$ and
\[
    \sum_{e\in V_{\mathrm{now}}}z_e(p)
    =
    \sum_{i=0}^m p_i|O_i|
    \le\kappa.
\]
Thus the type simplex parametrizes
$\operatorname{conv}\{\dTwoOne_{O_0},\ldots,\dTwoOne_{O_m}\}
    \subseteq\mathcal P_\kappa$.
The pure type $\delta_i$ maps to the core $O_i$, and every affine
segment in type space maps to an affine segment of feasible
physical vectors. In particular, a polygonal path in $\Delta$
gives a feasible fractional path with at most the same number
of segments.

Feasibility alone, however, does not ensure that a path maintains
enough value: the high-value regions of $F_Q$ need not be convex.
To certify value along our paths, we use the objective $H_j$
of~\eqref{eq:H-function}. Recall that $I(p)\subseteq\{0,\ldots,m\}$ is the
random subset obtained by including each type $i$ independently with
probability $p_i$. For a fixed buffer $Q$, write $H_Q(p)$ for the same
objective with $R_j$ replaced by $Q$; thus $H_{R_j}=H_j$.
Even though $\sum_i p_i=1$, the random set $I(p)$ may contain
zero, one, or several types. Consequently, the sampled union
need not satisfy the cardinality constraint, and $H_j(p)$ is
generally different from the linear average
$\sum_i p_i f(R_j\cup O_i)$. The independent samples defining
$H_j$ and $F_Q$ are analytical devices; their deterministic
realization is handled later.

The advantage of $H_j$ is that it depends only on values of
unions of the candidates, giving explicit formulas when
few types are involved. Submodularity makes it a valid
lower bound on the physical fractional objective and, as stated in
Section~\ref{sec:det-framework}, at most the linear average.
The same formula defines $H_Q(p)$ on the whole \emph{type cube}
$[0,1]^{m+1}$, not only on the simplex $\Delta$; we need this because the
rounding in Appendix~\ref{d2:app:rounding} also evaluates $H_Q$ at
vectors of mass below one.

\begin{dettwolemma}[Mixture comparisons]\label{d2:lem:lift}
    Fix a buffer $Q$. For every $p\in[0,1]^{m+1}$ with $\sum_i p_i\le1$,
    \begin{equation}\label{d2:eq:lift}
        F_Q(z(p))\ge H_Q(p),
    \end{equation}
    and for every mixture $p\in\Delta$,
    \begin{equation}\label{d2:eq:average}
        H_Q(p)\le\sum_i p_i f(Q\cup O_i).
    \end{equation}
\end{dettwolemma}
\begin{proof}
    For~\eqref{d2:eq:average}, subadditivity after contracting $Q$ gives
    \[
        f\Bigl(Q\cup\bigcup_{i\in I}O_i\Bigr)-f(Q)
        \le\sum_{i\in I}\bigl(f(Q\cup O_i)-f(Q)\bigr).
    \]
    Taking expectations and using $\sum_i p_i=1$ proves the claim.

    To prove~\eqref{d2:eq:lift}, define, on the full type cube
    $u\in[0,1]^{m+1}$,
    \[
        \Phi(u)
        =
        F_Q\!\left(
        \min\left\{\dTwoOne,\sum_{i=0}^m u_i\dTwoOne_{O_i}\right\}
        \right),
    \]
    where the minimum is coordinatewise.
    The multilinear extension of a monotone submodular function
    has nonnegative first derivatives and nonpositive mixed
    second derivatives, and hence is concave along every
    nonnegative direction.
    Fixing all but one coordinate of $u$, between successive
    capping points, the physical vector moves in a nonnegative
    direction, so $\Phi$ is concave there. At a capping point,
    the derivative loses only nonnegative contributions.
    Therefore $\Phi$ is concave in each type coordinate separately.

    Apply the chord inequality successively in all type
    coordinates. If $B_i$ are independent Bernoulli variables
    with means $p_i$, this gives
    \[
        \Phi(p)\ge\dTwoE[\Phi(B)].
    \]
    At a Boolean vector $B$, the capped physical vector is
    exactly the indicator of
    $\bigcup_{i:B_i=1}O_i$, so the right-hand side is $H_Q(p)$.
    Since $\sum_i p_i\le1$, no capping is needed at $p$,
    and $\Phi(p)=F_Q(z(p))$. This proves~\eqref{d2:eq:lift}.
\end{proof}

At a pure type, these quantities agree:
\[
    F_Q(z(\delta_i))=H_Q(\delta_i)=f(Q\cup O_i).
\]

We can therefore construct fractional paths by working
entirely in the type simplex: every value guarantee proved
for $H_j$ transfers to the corresponding physical vector.

\paragraph{Two-type interpolation.}
Before giving the reduction and the elementary certificate
construction, we record a local estimate for mixtures supported on
two types. It serves two purposes: protecting the bridges, that is, the connecting segments that join
certificates for consecutive windows, and bounding the value of
the half-mixture anchor in the elementary construction.

Fix a buffer $Q$ and two cores $A,B$. For $s\in[0,1]$, let
$H_Q(A,B;s)$ denote the type objective obtained by selecting
$A$ with probability $1-s$ and $B$ with probability $s$,
independently. Thus $s=0$ represents the pure core $A$,
$s=1$ represents the pure core $B$, and varying $s$ traverses
the edge between their type vectors.

Write
\[
    \lambda=f(Q),\qquad
    a=f(Q\cup A)-\lambda,\qquad
    b=f(Q\cup B)-\lambda,
\]
and define their residual overlap by
\[
    o=f(Q\cup A)+f(Q\cup B)-f(Q\cup A\cup B)-\lambda.
\]
Equivalently,
$f(Q\cup A\cup B)=\lambda+a+b-o$:
the term $o$ measures how much the sum of the two residual
values overcounts their combined contribution.
Submodularity gives $o\ge0$, and monotonicity gives
$o\le\min(a,b)$.

\begin{dettwolemma}[Two-type interpolation]\label{d2:lem:two-type}
    For every $s\in[0,1]$,
    \begin{equation}\label{d2:eq:edge}
        H_Q(A,B;s)
        =
        \lambda+(1-s)a+sb-s(1-s)o.
    \end{equation}
    Consequently, at the midpoint,
    \begin{equation}\label{d2:eq:half-lower}
        H_Q(A,B;1/2)
        \ge
        \lambda+\tfrac12(a+b)-\tfrac14\min(a,b).
    \end{equation}
    Moreover, if both cores have value at least $W$ under
    the same fixed buffer $Q$, that is,
    \[
        f(Q\cup A)\ge W,
        \qquad
        f(Q\cup B)\ge W,
    \]
    then every point along the edge has value at least $3W/4$.
\end{dettwolemma}
\begin{proof}
    The events that neither type, only $A$, only $B$,
    or both types are selected have probabilities
    $s(1-s)$, $(1-s)^2$, $s^2$, and $s(1-s)$, respectively.
    Averaging their values gives~\eqref{d2:eq:edge}.
    Setting $s=1/2$ and using $o\le\min(a,b)$ gives~\eqref{d2:eq:half-lower}.

    For the last claim, put $v=\min(a,b)$. Since
    $o\le v$ and $s(1-s)\le1/4$,
    \[
        \begin{aligned}
            H_Q(A,B;s)
             & \ge \lambda+\bigl(1-s(1-s)\bigr)v      \\
             & \ge \lambda+\tfrac34v                  \\
             & =\tfrac34\min\{f(Q\cup A),f(Q\cup B)\}
            +\tfrac14\lambda                          \\
             & \ge\tfrac34W.
        \end{aligned}
    \]
\end{proof}

Formula~\eqref{d2:eq:edge} expresses the value along the edge as the
linear interpolation of the endpoint values, less an
overlap penalty.
Inequality~\eqref{d2:eq:half-lower} bounds the value of a half-mixture anchor using
only the buffer value and the two endpoint residuals.
Applying such a bound at each buffer stage will help
establish persistence.

The factor $3/4$ is sharp for this interpolation:
when $\lambda=0$ and $a=b=o=W$, the midpoint has
value exactly $3W/4$.

The last claim of the lemma provides a safe bridge between
two cores that meet a common benchmark under one objective.
In the reduction, we use it to join certificates
for consecutive windows. By~\eqref{d2:eq:lift}, both the
anchor and path estimates also hold for the corresponding
physical fractional vectors. The remaining geometric task
is to choose an anchor and paths that satisfy the
certificate requirements across the entire window.

\subsubsection{An elementary certificate at level \texorpdfstring{$1/\sqrt2$}{1/sqrt(2)}}
\label{d2:sec:half}

We prove Theorem~\ref{thm:det_sqrt2} through a more general
prefix--suffix tradeoff. Recall that the anchor is the half-mixture
$p=\tfrac12(\delta_0+\delta_c)$. The partner $O_c$ is an optimizer at a
carefully chosen future stage. The selection rule balances the constraints
before and after that stage; the same estimates also certify the entry
and exit paths.

Fix $A=O_0$ and define
\begin{equation}\label{d2:eq:K}
    r_j=f(R_j\cup A)-\lambda_j,\qquad K_j=2g_j-r_j.
\end{equation}
Here $g_j$ is the remaining optimum and $g_j-r_j$ is the initial
core's deficit; $K_j$ is their sum. We have $0\le r_j\le g_j$, and
\begin{equation}\label{d2:eq:K-bound}
    K_j\le P_j.
\end{equation}
To see this, note that Lemma~\ref{d2:lem:window-bounds} and monotonicity give $g_j\le V\le f(R_j\cup A)=\lambda_j+r_j$;
therefore $2g_j-r_j\le\lambda_j+g_j=P_j$.

\begin{dettwotheorem}[Prefix--suffix tradeoff]\label{d2:thm:half}
    For $1/2<\beta\le3/4$, set
    \begin{equation}\label{d2:eq:selection}
        \gamma=4\beta-2,\qquad
        D=\max_{0\le j\le m}\{2K_j-4(1-\beta)P_j\},\qquad
        c=\max\{j:K_j\ge D\}.
    \end{equation}
    The index $c$ exists. The half-mixture of $O_0$ and $O_c$ has value
    at least $\beta P_j$ for $j\le c$ and at least $P_j/(2\beta)$ for
    $j>c$. It admits an entry path of one segment and an exit path of at
    most two segments at common level
    \begin{equation}\label{d2:eq:tradeoff}
        \alpha(\beta)=\min\{\beta,1/(2\beta)\}.
    \end{equation}
\end{dettwotheorem}

The envelope in~\eqref{d2:eq:selection} records the contribution
needed from the partner to protect the prefix. Choosing the last
eligible partner turns every later failure of eligibility into a
suffix guarantee. We establish the two inequalities behind this
argument before proving the theorem.

\begin{dettwolemma}[Prefix bound]\label{d2:lem:prefix}
    If $B=O_c$ is the candidate at stage $c$, then at every
    stage $j\le c$,
    \begin{equation}\label{d2:eq:prefix}
        H_{R_j}(A,B;1/2)\ge\lambda_j+\tfrac12r_j+\tfrac14K_c.
    \end{equation}
\end{dettwolemma}
\begin{proof}
    Diminishing returns gives
    \[
        f(R_j\cup B)-\lambda_j\ge g_c
    \]
    and
    \begin{align*}
        f(R_j\cup A\cup B)-f(R_j\cup A)
         & \ge f(R_c\cup A\cup B)-f(R_c\cup A)             \\
         & \ge (f(R_c\cup B)-f(R_c))-(f(R_c\cup A)-f(R_c)) \\ 
         & \ge g_c-r_c.
    \end{align*}
    The sum of the values of the four equally likely type outcomes is
    therefore at least
    \[
        \lambda_j+(\lambda_j+r_j)+(\lambda_j+g_c)
        +(\lambda_j+r_j+g_c-r_c)
        =4\lambda_j+2r_j+K_c.
    \]
    Divide by four.
\end{proof}

\begin{dettwolemma}[Suffix bound]\label{d2:lem:suffix}
    Let $\lambda,g,u,z\ge0$ and $0\le\gamma\le2$. If
    $2g-u\le\gamma(\lambda+z)$, then
    \begin{equation}\label{d2:eq:suffix}
        \lambda+\tfrac12(u+z)-\tfrac14\min(u,z)
        \ge\frac{2}{2+\gamma}(\lambda+g).
    \end{equation}
\end{dettwolemma}
\begin{proof}
    The premise gives
    $2(\lambda+g)\le(2+\gamma)\lambda+u+\gamma z$.
    Put $w=\min(u,z)$. Since the coefficients of $u,z$ below are
    nonnegative, we get:
    \begin{align*}
         & (2+\gamma)\left[\tfrac12(u+z)-\tfrac14w\right]-u-\gamma z \\
         & \quad=\tfrac\gamma2u+(1-\tfrac\gamma2)z
        -\tfrac{2+\gamma}{4}w                                        \\
         & \quad\ge\left(\tfrac\gamma2+1-\tfrac\gamma2
        -\tfrac{2+\gamma}{4}\right)w
        =\tfrac{2-\gamma}{4}w\ge0.
    \end{align*}
    Adding $(2+\gamma)\lambda$ and dividing by $(2+\gamma)$ proves the claim as follows.
    \begin{align*}
         & (2+\gamma)\left[\tfrac12(u+z)-\tfrac14w\right]-u-\gamma z \ge 0 \implies                                             \\
         & (2+\gamma)\left[\lambda + \tfrac12(u+z)-\tfrac14w\right]-u-\gamma z \ge (2+\gamma)\lambda \implies                   \\
         & (2+\gamma)\left[\lambda + \tfrac12(u+z)-\tfrac14w\right] \ge (2+\gamma)\lambda +u+\gamma z \ge 2(\lambda+g) \implies \\
         & \lambda + \tfrac12(u+z)-\tfrac14w \ge \frac{2}{2+\gamma}(\lambda+g)
    \end{align*}
\end{proof}

The prefix bound explains the envelope in Theorem~\ref{d2:thm:half}, particularly why we define variable $D$ and index $c$ in this way.
To obtain level $\beta$ at stage $j$, we should show:
\[
    \lambda_j+\tfrac12r_j+\tfrac14K_c \ge \beta P_j\iff
    K_c \ge 4\beta P_j -4\lambda_j -2 r_j
\]
We rewrite this requirement as follows:
\begin{align*}
    K_c & \ge 4\beta P_j -4\lambda_j -2 r_j             \\
        & = 4 P_j -4\lambda_j -2 r_j - 4(1 - \beta) P_j \\
        & = 4 g_j -2 r_j - 4(1 - \beta) P_j             \\
        & = 2K_j - 4(1 - \beta) P_j                     \\
\end{align*}
Therefore to obtain level $\beta$ at stage $j$, it is sufficient that
$K_c \ge 2K_j - 4(1 - \beta) P_j$.

To obtain level $\beta$ at stage $j$, it is sufficient that
\[
    K_c\ge4\beta P_j-4\lambda_j-2r_j
    =2K_j-4(1-\beta)P_j.
\]
Taking the maximum of the right side gives a single requirement on
the partner.
In Theorem~\ref{d2:thm:half}, we choose index $c$ as the last partner that meets this requirement. This will ensure level $\beta$ is obtained up to index $c$. The maximality in definition of $c$
will turn every later failure of eligibility into the premise of Lemma~\ref{d2:lem:suffix} (Suffix bound lemma).

\begin{proof}[Proof of Theorem~\ref{d2:thm:half}]
    \emph{Existence of the partner and its benchmark credit.}
    At stage zero $K_0=P_0=V$, so $D\ge\gamma V>0$.
    If index $h$ attains the maximum value of expression that defines $D$ in~\eqref{d2:eq:selection}, then~\eqref{d2:eq:K-bound} gives
    \begin{align*}
        D & =2K_h-4(1-\beta)P_h\le(4\beta-2)K_h=\gamma K_h\le K_h, \\
        D & \le(4\beta-2)P_h=\gamma P_h.
    \end{align*}
    Here $0<\gamma\le1$ because $\beta \le 3/4$.
    The first inequality, $D \le K_h$, implies that index $h$ is an eligible choice. In other words, $h \in \{j: K_j \ge D\}$. This ensures the set of eligible candidates is non-empty, and therefore index $c$ is well-defined (it exists).
    Since $h$ is eligible, $c\ge h$, and Lemma~\ref{d2:lem:window-bounds} gives
    \begin{equation}\label{d2:eq:credit}
        P_c\ge P_h\ge D/\gamma.
    \end{equation}
    Let $B=O_c$.

    \emph{Persistence on the prefix.}
    For $j\le c$, Lemma~\ref{d2:lem:prefix} and the definition of $D$ yield
    \begin{align*}
        4\bigl(H_{R_j}(A,B;1/2)-\beta P_j\bigr)
         & \ge K_c-\bigl[2K_j-4(1-\beta)P_j\bigr] \\
         & \ge K_c-D\ge0.
    \end{align*}

    \emph{Persistence on the suffix.}
    For $j>c$, put $u=r_j$ and $z=f(R_j\cup B)-\lambda_j$.
    Last eligibility, i.e. $K_j < D$, and~\eqref{d2:eq:credit} imply
    \begin{equation}\label{d2:eq:suffix-premise}
        2g_j-u=K_j<D\le\gamma P_c\le\gamma(\lambda_j+z).
    \end{equation}
    The last inequality uses monotonicity of $f$, namely $f(R_j\cup B) \ge f(R_c \cup B)$, and the
    optimality of $B$ at $c$. Lemma~\ref{d2:lem:suffix}, together with
    \eqref{d2:eq:half-lower}, gives
    \begin{equation}\label{d2:eq:reserve}
        H_{R_j}(A,B;1/2)\ge
        L_j:=\lambda_j+\tfrac12(u+z)-\tfrac14\min(u,z)
        \ge\frac{2}{2+\gamma}P_j=\frac{P_j}{2\beta}.
    \end{equation}

    \emph{Safe entry.}
    Use the single edge from $A$ to its midpoint with $B$. Initially $f(B)\le
        f(A)=V$, and differentiation of~\eqref{d2:eq:edge} gives
    \[
        \frac{d}{ds}H_{\varnothing}(A,B;s)
        =f(B)-V-(1-2s)o\le0\qquad(0\le s\le1/2).
    \]
    Since the derivative is non-positive throughout the entry, from $s:0 \rightarrow 1/2$, the entire entry is at least its final value, which is at least
    $\beta V$ by the prefix calculation (Lemma~\ref{d2:lem:prefix}).

    \emph{Safe exit.}
    We start by lower bounding the value at the starting point of the exit path. We would like to show $H_m(p)\ge\alpha P_m$, where $\alpha=\alpha(\beta)$.
    At the terminal stage write $u=r_m$, $z=f(R_m\cup B)-\lambda_m$.
    If $c<m$,~\eqref{d2:eq:reserve} gives $L_m\ge P_m/(2\beta)$.
    If $c=m$, then $z=g_m\ge u$, so
    \[
        L_m=\lambda_m+\tfrac14r_m+\tfrac12g_m
        =\lambda_m+\tfrac12r_m+\tfrac14K_m
        \ge\beta P_m
    \]
    by the same prefix calculation.
    We note that since $c=m$, the above inequality is equivalent to showing $K_c \ge 2K_c - 4(1-\beta)P_c$ which is achieved by definition of $c$ and $D$ in this case.
    Hence $L_m\ge\alpha P_m$ in both cases.

    From the midpoint move toward whichever of $A,B$ has larger terminal
    residual $v=\max(u,z)$. The value is nondecreasing on that half edge
    because throughout the movement toward the larger endpoint, the derivative of the value for is the nonnegative endpoint difference plus $(2s-1)o$. Note that $(2s-1)o$ is also non-negative since $s\in[1/2,1]$.
    So far, we are maintaining the lower bound value of $\alpha P_m$ in our exit path.
    Next move from that endpoint
    (whichever of $A,B$ has larger terminal residual $v=\max(u,z)$)
    to the prescribed terminal optimizer
    $O_m$, whose residual is $g_m\ge v$. The second edge has value at
    least
    \[
        \lambda_m+[1-s(1-s)]v\ge\lambda_m+\tfrac34v
        \ge\lambda_m+\tfrac12(u+z)-\tfrac14\min(u,z)=L_m \ge \alpha P_m.
    \]
    Both exit segments therefore satisfy the required terminal guarantee.

    If $c=0$, perform the calculation using a second label for the same
    core $O_0$. Merge the two labels afterwards. For duplicate types with
    probabilities $a,b$, separate selection includes their common set with
    probability $a+b-ab$, whereas the merged label includes it with
    probability $a+b$. Monotonicity shows that merging can only increase
    every value. It maps affine paths to affine paths and turns the
    anchor into $\delta_0$, so all the conclusions hold on the original
    simplex.

    Finally, $\beta$ increases and $1/(2\beta)$ decreases. Their balance
    is $\beta^2=1/2$, giving $1/(2\beta)=\beta=1/\sqrt2$. The physical
    lift~\eqref{d2:eq:lift} preserves the anchor and its paths.
\end{proof}

Taking $\beta=1/\sqrt2$ proves Theorem~\ref{thm:det_sqrt2}, with one
entry segment and at most two exit segments.

The limiting ratio $1/\sqrt2\approx0.707107$ exceeds
$1-1/e\approx0.632121$. Its source is the balance between prefix and
suffix persistence. The transition analysis preserves that balance:
the suffix calculation gives the stronger reserve $L_m$ needed for
the exit, and the entry is monotone toward the anchor. The rational choice $\beta=7/10$ gives prefix level $7/10$ and suffix
level $5/7$, hence a $0.7-\eps$ guarantee by the same proof.

\subsubsection{Simultaneous deterministic rounding}
\label{d2:app:rounding}
\label{d2:sec:transfer}

We now prove Theorem~\ref{thm:det_reduction}; the proof continues in
Appendix~\ref{d2:app:schedule}.
We first prove the simultaneous rounding statement. Fix a certificate at level
$\alpha\le3/4$ with $L$ segments. All its cores are subsets of the
same set $V_{\mathrm{now}}$, have size at most $\kappa$, and have value at most $V$.
Fix $0<\eta<1/100$.

\paragraph{Reducing number of stages to $m=O(1/\eta)$.}
We first show that we can always reduce the number of stages $j=0, 1, \cdots, m$ to $O(1/\eta)$ by losing at most $O(\eta)$ in the approximation guarantee.
We simply keep a subset of stages $j \in \{0, 1, \cdots, m\}$ as follows. We always keep the first and last stages $j=0$ and $j=m$. We keep track of the value of the buffer set $R_j$. For every non-negative integer $i$, we assign stage $j$ to bin $i$ if the value $f(R_j)$ is in range
$[i\eta V,(i+1)\eta V)$. From each bin with $i\eta \le 3$, we keep the first stage in it. Note that for larger value bins, the buffer set already provides a value of at least $3V$, while $P_j\le\lambda_j+V$ by Lemma~\ref{d2:lem:window-bounds}. In those stages, the approximation factor of $\nicefrac{3}{4}$ is guaranteed independent of what sets we choose for the certificate. Since the value of buffer set does not change by more than $\eta V$ in each bin, neither does the benchmark $P_j$ by Lemma~\ref{d2:lem:window-bounds}, so retaining only one stage suffices to have the same approximation guarantee with an extra loss of $\eta$. This reduction ensures that we can always work with only $m = O(1/\eta)$ stages.

\paragraph{Finite tests and continuity.}
Every type marginal under every buffer is at most $f(O_i)\le V$.
Consequently, on the type cube $[0,1]^{m+1}$,
\begin{equation}\label{d2:eq:lipschitz}
    |H_Q(p)-H_Q(q)|\le V\|p-q\|_1.
\end{equation}

Indeed, changing one type probability changes the expectation at a rate equal to that type's expected marginal contribution, which lies in $[0,V]$.

If needed, approximate the original path vertices and anchor by
rational simplex points within $\eta$ in $\ell_1$, retaining the
same shared anchor and keeping pure endpoints fixed.
For each path vertex, one can always find a simplex point with $\ell_1$ distance at most $\eta$ to the vertex and also its coordinates are all rational.
Linear interpolation stays within $\eta$ of each original segment.
To ensure segments are not too long, sub-divide each segment into sub-segments of length at most $\eta$ in $\ell_1$ distance. This is doable while keeping coordinates rational, and it will increase the number of segments/vertices by a factor of $1/\eta$.
Include value
tests at all sampled points, all retained anchor stages, and the
additional safety points described below; include cardinality tests
at the sampled points. The total number $M$ of tests is
\begin{equation}\label{d2:eq:test-count}
    M=O((L+1)/\eta).
\end{equation}
We may increase $M$ to be at least two. Each value comparison is to
its corresponding ideal point; the error does not accumulate along
the path. A value test at a rounded set simply checks that the realized value of the set does not decrease from its expected value by more than $O(\eta)$ fraction. The cardinality test checks its size does not increase by more than this fraction.

\paragraph{Force the elements with large marginal value.}
Let $U$ be the union of the cores used in the certificate. Set
\[
    \theta=\frac{\eta^2}{4\log(8M)},\qquad
    h=\lceil1/\theta\rceil.
\]
Assume $\kappa\ge h$. Starting with $D=\varnothing$, greedily add
an element of $U$ while its marginal under $f$ exceeds $\theta V$.
The procedure selects fewer than $h$ elements: otherwise its first
$h\le\kappa$ elements would have value greater than
$h\theta V\ge V$. Thus
\begin{equation}\label{d2:eq:heavy-size}
    |D|<h=O(\eta^{-2}\log M).
\end{equation}
After contracting $D$ and any future buffer, every remaining
singleton marginal is at most $\theta V$. Here the contraction means adding set $D$ and the buffer to all sets during the entire window $j=0, \cdots, m$.

\paragraph{Use one threshold per element.}
Draw independent uniforms $U_e$ for $e\in U\setminus D$, shared by
all sampled points, both paths, and the anchor. For a type
vector $p$ of mass at most one, set
\[
    T(p)=D\cup\{e\in U\setminus D:U_e\le z_e(p)\}.
\]
Forcing $D$ can only improve the expected value. Hence Lemma~\ref{d2:lem:lift} gives
$\dTwoE[f(Q\cup T(p))]\ge H_Q(p)$.

We use two standard concentration bounds.

\begin{dettwolemma}[Concentration]\label{d2:lem:concentration}
    Let $G$ be a nonnegative monotone submodular set function whose
    marginal values are at most $c>0$, let $S$ be a random set containing
    each element independently, and put $\mu=\dTwoE[G(S)]$. Then, for $0<a<\mu$,
    \begin{equation}\label{d2:eq:submod-tail}
        \dTwoPr\{G(S)\le\mu-a\}
        \le\exp\left(-\frac{a^2}{2c\mu}\right).
    \end{equation}
    If $X$ is a sum of independent Bernoulli variables with mean at
    most $\kappa$, then, for $0<\eta\le1$,
    \begin{equation}\label{d2:eq:chernoff}
        \dTwoPr\{X>(1+\eta)\kappa\}\le e^{-\eta^2\kappa/3}.
    \end{equation}
\end{dettwolemma}
\begin{proof}
    The first bound is \citet[Corollary~3.2]{Vondrak10}, applied to $G/c$
    with $\delta=a/\mu$. The second is the standard multiplicative Chernoff
    bound, applied with the upper bound $\kappa$ on the mean.
\end{proof}

To bound the deviation independently of $|U|$, put
$G_Q(S)=f(Q\cup D\cup S)-f(Q\cup D)$. Submodularity gives
$G_Q(S)\le f(S)$. Partitioning $S$ into sets of size at most
$\kappa$ and using subadditivity yields
\[
    G_Q(S)\le f(S)\le\left\lceil\frac{|S|}{\kappa}\right\rceil V
    \le\left(1+\frac{|S|}{\kappa}\right)V.
\]
The random part of $T(p)$ has expected size at most $\kappa$, so
$\dTwoE[G_Q] \le2V$. After contracting $D$, the marginal values of $G_Q$ are
at most $\theta V$, and the random part of $T(p)$ contains each element
independently. Since $H_Q(p)\le \dTwoE[f(Q\cup T(p))] = f(Q\cup D) + \dTwoE[G_Q]$,
applying~\eqref{d2:eq:submod-tail} with $c=\theta V$ and $a=\eta V$ gives
\begin{equation}\label{d2:eq:value-failure}
    \dTwoPr\{f(Q\cup T(p))<H_Q(p)-\eta V\}
    \le\exp\{-\eta^2/(4\theta)\}=\frac1{8M}.
\end{equation}
If the residual mean is at most $\eta V$, the corresponding downward
deviation is impossible, so the same assertion holds. The size of the
random part of $T(p)$ is a sum of independent Bernoulli variables,
so~\eqref{d2:eq:chernoff} gives
\begin{equation}\label{d2:eq:size-failure}
    \dTwoPr\{|T(p)\setminus D|>(1+\eta)\kappa\}
    \le e^{-\eta^2\kappa/3}\le\frac1{8M},
\end{equation}
using $\kappa\ge h\ge4\eta^{-2}\log(8M)$. Note that for smaller values of $\kappa$ (and hence $k$), the consistency objective is achieved trivially since the change from any solution to the next is $2k$-consistent.

On an original segment with endpoints $p,q$, each physical
probability varies affinely. The expected number of membership
changes is
\[
    \sum_{e\notin D}|z_e(p)-z_e(q)|
    \le\sum_i |p_i-q_i|\,|O_i|
    \le\kappa\|p-q\|_1\le2\kappa.
\]
Thus the expected number on both paths is at most $2L\kappa$.
Markov's inequality shows that it exceeds $8L\kappa$ with
probability at most $1/4$. Together with all the value and
cardinality tests, the union bound has probability below one.
Fix one successful threshold assignment deterministically. With
rational sampled probabilities there are finitely many threshold
cells, so exhaustive search suffices. Rational certificates with the
allowed slack can likewise be found by enumerating rational path
vertices and checking a sufficiently fine mesh using
\eqref{d2:eq:lipschitz}. No efficiency bound is claimed.

\paragraph{Protect every intermediate set.}
For adjacent subdivisions $p,q$, put $v_i=\min(p_i,q_i)$. This is a
type vector of mass at most one, and $\|p-v\|_1\le\eta$. Include
the value test for $T(v)$ under the objective used on that path.
Since $z_e(v)\le\min(z_e(p),z_e(q))$, common thresholds ensure
\[
    T(v)\subseteq T(p)\cap T(q).
\]
Delete $T(p)\setminus T(q)$, then insert $T(q)\setminus T(p)$.
Every intermediate set contains $T(v)$ and has size at most the
larger endpoint size. The Lipschitz loss to $v$, the rationalization
loss, and the concentration loss are together $O(\eta)V$.

At a pure type $\delta_i$, $T(\delta_i)=O_i\cup D$. Add $D$ at the initial
pure endpoint and remove it at the terminal pure endpoint; these
extra operations preserve the endpoint value. We obtain a single
rounded anchor with coupled entry and exit, such that
\begin{align}
    \text{core size}        & \le(1+\eta)\kappa+h,
    \label{d2:eq:rounded-size}                     \\
    \text{total operations} & \le8L\kappa+2h.
    \label{d2:eq:rounded-operations}
\end{align}
The anchor is $(\alpha-O(\eta))$-good at every buffer stage,
and every entry and exit intermediate set loses at most $O(\eta)V$
from its required value. A one-edge bridge between pure cores is
rounded by the same argument, using its own forced set and thresholds.
The estimates remain valid when the fixed buffer overlaps a core: its
elements then have zero marginal value after contraction.
The pure endpoints allow these forced sets to be removed before the
next is added.

\subsubsection{The protected schedule and the recourse calculation}
\label{d2:app:schedule}

We complete the proof of Theorem~\ref{thm:det_reduction}. Write $X_t$ for the
arrived prefix and
\[
    \OPT_t(r)=\max\{f(A):A\subseteq X_t,\ |A|\le r\},
\]
so that $\OPT_t(k)=f(\OPT_t)$. First, for $1\le r\le k$,
\begin{equation}\label{d2:eq:thinning}
    \OPT_t(r)\ge\frac r k\OPT_t(k).
\end{equation}
Indeed, for a normalized monotone submodular function and a set $S$,
diminishing returns and telescoping imply
\[
    \sum_{e\in S}\bigl[f(S)-f(S\setminus\{e\})\bigr]\le f(S).
\]
Some deletion therefore keeps at least $(|S|-1)/|S|$ of the value.
Repeated deletion from an optimal set of size at most $k$ proves
\eqref{d2:eq:thinning}.

\paragraph{Parameters and transition intervals.}
Fix sufficiently small $\eta>0$ and put
\[
    \kappa=\lfloor(1-10\eta)k\rfloor,\quad
    b=\lfloor\eta k\rfloor,\quad J=\lceil4/\eta\rceil,\quad
    \ell=\lfloor b/J\rfloor.
\]
For now assume
$k\ge K_\eta=c\eta^{-3}\log(e/\eta)$ for a sufficiently large
constant $c$. Then $\kappa\ge1$ and $\ell=\Theta(\eta^2k)$.
Keep the rolling buffer
\[
    R_t=X_t\setminus X_{\max\{0,t-4b\}},
\]
which uses at most $4b$ slots and changes by at most two elements
per update.

For each block starting at $b_i=ib$, $i\ge1$, inspect the $J$
successive intervals $[b_i+j\ell,b_i+(j+1)\ell]$,
$0\le j<J$. Choose the first interval $[u,u+\ell]$ for which
\begin{equation}\label{d2:eq:stable}
    \OPT_{u+\ell}(k)\le(1+\eta)\OPT_u(k).
\end{equation}
If none exists, choose the last interval. Let $\tau_i$ be the
chosen start. The intervals lie within their blocks and do not
overlap in their interiors.

When no interval is stable, the preceding $J-1$ increases give
\[
    \OPT_{\tau_i}(k)\ge(1+\eta)^{J-1}\OPT_{b_i}(k)
    \ge4\OPT_{b_i}(k).
\]
For the last inequality, Bernoulli's inequality gives
$(1+\eta)^{J-1}\ge1+(J-1)\eta\ge5-\eta>4$.
For every $t\in[\tau_i,\tau_i+\ell]$, let $B_t=X_t\setminus X_{b_i}$.
Split an optimal solution into its old part and its part in $B_t$.
Subadditivity and monotonicity imply
\begin{equation}\label{d2:eq:buffer-protection}
    f(B_t)\ge\OPT_t(k)-\OPT_{b_i}(k)\ge\tfrac34\OPT_t(k).
\end{equation}
Moreover $B_t\subseteq R_t$. Thus each chosen interval is protected
either by~\eqref{d2:eq:stable} or by the buffer itself. This includes
the case $\OPT_{b_i}(k)=0$.

\paragraph{The local windows.}
Window $i$ freezes $N_i=X_{\tau_i}$ as its set $V_{\mathrm{now}}$ and uses the growing buffer
$R_i(t)=X_t\setminus X_{\tau_i}$ for
$\tau_i\le t\le\tau_{i+1}$. Its initial value is
$V_i=\OPT_{\tau_i}(\kappa)$. Its local benchmark satisfies
\begin{equation}\label{d2:eq:local-global}
    P_i(t)\ge\OPT_t(\kappa),
\end{equation}
because the old part of any size-$\kappa$ solution is a feasible
old core and all of its new elements lie in $R_i(t)$.

Choose a prescribed initial optimizer $O_i$ of value $V_i$ and a
prescribed terminal old-core optimizer $C_i$ for the buffer at
$\tau_{i+1}$; these are the candidates $O_0$ and $O_m$ of window $i$. Retain $O(1/\eta)$ stages and apply the fractional
hypothesis, then Appendix~\ref{d2:app:rounding}. Denote the rounded
anchor by $T_i$. After transition $i$, hold $T_i$ until
transition $i+1$. If $V_i=0$, take the old core empty: all its
residuals are zero and the local buffer gives the full benchmark.

\paragraph{Joining two windows.}
During transition $i$, for $i\ge2$, execute the following ordered
list: the old exit from $T_{i-1}$ to $C_{i-1}$; a bridge from
$C_{i-1}$ to $O_i$; the new entry from $O_i$ to $T_i$. The old
exit is evaluated under the fixed buffer
$Q=R_{i-1}(\tau_i)$ and gives
\[
    \alpha P_{i-1}(\tau_i)-O(\eta)V_{i-1}
    \ge(\alpha-O(\eta))V_i,
\]
since $P_{i-1}(\tau_i)\ge V_i$ and $V_{i-1}\le V_i$.
Both bridge endpoints, under this same $Q$, have value at least
$V_i$. Lemma~\ref{d2:lem:two-type} therefore
gives $3V_i/4$ before rounding. Round this bridge in old set $N_i$,
with normalization $V_i$; it loses $O(\eta)V_i$. Finally the new
entry gives $(\alpha-O(\eta))V_i$ even with an empty buffer.
All elements used have arrived by $\tau_i$.

At the shared pure cores, remove the previous forced set before
adding the next. During this operation the pure core remains
present, so the endpoint value is preserved and the forced sets
do not accumulate. A zero-value old window needs no rounding; if
the new value is positive, its old terminal endpoint is still
protected by $Q$ and the same bridge argument applies. If the
new value is also zero, the newly arriving buffer supplies all
value during the transition.

The required buffer $Q$ remains inside $R_t$ until the transition
ends: $\tau_i+\ell-\tau_{i-1}\le2b$. The same age bound protects
$R_i(t)$ throughout the subsequent holding interval. Additional
arrivals can only improve the values of the protected sets.

On a stable transition interval,~\eqref{d2:eq:thinning} and
\eqref{d2:eq:stable} give
\begin{align*}
    f(\ALG_t) & \ge(\alpha-O(\eta))V_i                               \\
              & \ge(\alpha-O(\eta))\frac{\kappa}{k(1+\eta)}\OPT_t(k)
    =(\alpha-O(\eta))\OPT_t(k).
\end{align*}
On an unstable transition interval,~\eqref{d2:eq:buffer-protection}
gives the stronger $3/4$ guarantee. During holding intervals,
anchor persistence,~\eqref{d2:eq:local-global}, and thinning
give the same $(\alpha-O(\eta))$ ratio directly.

\paragraph{Capacity, worst-case changes, and endpoints.}
After retaining the stages, $L=O(1/\eta)$ and the rounding test
count is $M=O(\eta^{-2})$. Increasing $c$ in $K_\eta$ ensures that
every forced set has size at most $\eta k$ and that all rounding
premises hold. Equations~\eqref{d2:eq:rounded-size} and the buffer
size give
\[
    |\ALG_t|\le(1+\eta)\kappa+\eta k+4b\le k.
\]
Deletion-first updates preserve the same bound at every
intermediate operation. The concatenated list at a transition
contains
\[
    O(L\kappa+\eta^{-2}\log(e/\eta))=O(k/\eta)
\]
operations. Execute at most the ceiling of this count divided by
$\ell$ at each of the $\ell$ actual updates. This gives
$O(\eta^{-3})$ operations per update, plus at most two buffer
changes. Full symmetric difference is at most the number of
insertions and deletions performed.

Before the first transition ends there are at most $2b$ arrived
elements, and the rolling buffer contains the entire prefix.
This protects initialization: construct $O_1$ and then follow its
rounded entry during that interval, adding at most $\kappa$ extra
operations. For a finite stream, append enough formal zero-value
arrivals to define the final windows and transitions, then
restrict the sequence to the actual stream. Every core used
before the real end consists of elements already present at its
window start; the padding does not authorize any unarrived real
element.

For $k<K_\eta$, recompute an exact optimum at each update, with
recourse at most $2k=O(\eta^{-3}\log(e/\eta))$. The constant consistency in this case is implied due to the constant upper bound on $k$. Choose $\eta$ a
sufficiently small multiple of $\eps$. For $\eps<\alpha\le3/4$,
$\log(e/\eps)=O(\log(1/\eps))$, as used in the theorem, and for
$\eps\ge\alpha$, the empty solution already gives the claimed nonpositive
approximation guarantee with zero recourse. This proves
Theorem~\ref{thm:det_reduction}. If $f(\varnothing)>0$, apply the
argument to $f-f(\varnothing)$; restoring the baseline only improves
an approximation ratio at most one.

\end{document}